\documentclass{aa}

\DeclareRobustCommand{\TUSSEN}[3]{#2}
\usepackage{graphicx}
\usepackage{txfonts}
\usepackage{color}

\begin{document}

\title{An interference detection strategy for Apertif\\based on AOFlagger~3}

\author{A. R. Offringa\inst{1,2}
  \and
  B. Adebahr\inst{3}
  \and
  A. Kutkin\inst{1}
  \and
  E.~A.~K.~Adams\inst{1,2}
  \and
  T.~A.~Oosterloo\inst{1,2}
  \and
  J.~M.~van~der~Hulst\inst{2}
  \and
  H.~Dénes\inst{1}
  \and
  C.~G.~Bassa\inst{1}
  \and
  D.~L.~Lucero\inst{4,1}
  \and
  % builders from here, alphabetically
  W.~J.~G.~Blok\inst{1,5,2}
  \and
  K.~M.~Hess\inst{6,1,2}
  \and
  J.~van~Leeuwen\inst{1}
  \and
  G.~M.~Loose\inst{1}
  \and
  Y.~Maan\inst{7,1}
  \and
  L.~C.~Oostrum\inst{1,8,9}
  \and
  E.~Orr\'u\inst{1}
  \and
  D.~Vohl\inst{8,1}
  \and
  J.~Ziemke\inst{1,10}
  }

\institute{ASTRON, the Netherlands Institute for Radio Astronomy, Oude Hoogeveensedijk 4,7991 PD Dwingeloo, The Netherlands\\
    \email{offringa@astron.nl}
  \and
    Kapteyn Astronomical Institute, University of Groningen, PO Box 800, 9700 AV Groningen, The Netherlands
  \and
  Astronomisches Institut der Ruhr-Universit{\"a}t Bochum (AIRUB), Universit{\"a}tsstrasse 150, 44780 Bochum, Germany
  \and
  Department of Physics, Virginia Polytechnic Institute and State University, 50 West Campus Drive, Blacksburg, VA 24061, USA
  \and
  Dept.\ of Astronomy, Univ.\ of Cape Town, Private Bag X3, Rondebosch 7701, South Africa
  \and
  Instituto de Astrofísica de Andalucía (CSIC), Glorieta de la Astronomía s/n, 18008 Granada, Spain
  \and
  National Centre for Radio Astrophysics, Tata Institute of Fundamental Research, Pune 411007, Maharashtra, India
  \and
  Anton Pannekoek Institute, University of Amsterdam, Postbus 94249, 1090 GE Amsterdam, The Netherlands
  \and
  Netherlands eScience Center, Science Park 402, 1098 XH Amsterdam, The Netherlands
  \and
  University of Oslo Center for Information Technology, P.O. Box 1059, 0316 Oslo, Norway
    }

\date{Received September 20, 2022; accepted January 3, 2023}

  \abstract
   {Apertif is a multi-beam receiver system for the Westerbork Synthesis Radio Telescope that operates at 1.1-1.5 GHz, which overlaps with various radio services, resulting in contamination of astronomical signals with radio-frequency interference (RFI).}
   {We analyze approaches to mitigate Apertif interference and design an automated detection procedure for its imaging mode. Using this approach, we present long-term RFI detection results of over 300 Apertif observations.}
   {Our approach is based on the AOFlagger detection approach. We introduce several new features, including ways to deal with ranges of invalid data (e.g. caused by shadowing) in both the SumThreshold and scale-invariant rank operator steps; pre-calibration bandpass calibration; auto-correlation flagging; and HI flagging avoidance. These methods are implemented in a new framework that uses the Lua language for scripting, which is new in AOFlagger version 3.}
   {Our approach removes RFI fully automatically, and is robust and effective enough for further calibration and (continuum) imaging of these data. Analysis of 304 observations show an average of 11.1\% of lost data due to RFI with a large spread. We observe 14.6\% RFI in auto-correlations. Computationally, AOFlagger achieves a throughput of 370 MB/s on a single computing node. Compared to published machine learning results, the method is one to two orders of magnitude faster.}
   {}


   \maketitle
%
%-------------------------------------------------------------------

\section{Introduction}

Technical advancement of mankind is driving an increase of man-made radio-frequency transmitters, both terrestrial and in space. This raises the bar for radio astronomical studies that try to detect sky signals that are many orders of magnitude fainter than man-made transmissions. Now that radio-astronomy is evolving into a science where it is the norm to measure data volumes in petabytes, mitigation of radio-frequency interference (RFI) needs to be computationally efficient and fully automated.

Apertif is a receiver system upgrade for the Westerbork Synthesis Radio Telescope (WSRT) that makes use of phased-array feeds to allow for 40 simultaneous adjacent beams on the sky \citep{apertif-van-cappellen-2022}. Observations are performed at a central frequency of 1280 or 1370~MHz with an instantaneous bandwidth of 300~MHz.

The data volume produced by Apertif is considerable. Voltages from the 12 dishes with Apertif receivers are correlated for all beams, typically integrated for 30 seconds and recorded with four polarizations. The bandwidth of 300~MHz is split into 384 sub-bands, each with 64 channels of 12.2~kHz. Because of the large bandwidth, it overlaps with various services, including GPS and air-traffic communications. Although the WSRT resides in a radio protected zone, it is not shielded from satellites and air-traffic. Moreover, starting 2020, 5G transmissions make use of the 1452 -- 1492 MHz bandwidth. For these reasons, Apertif requires an efficient approach to deal with RFI. Due to the large amount of data, such an approach has to work fully automatically.

The most common method to deal with RFI, is to detect data samples that have a significant contribution of RFI and ignore affected data in the processing (e.g. \citealt{statistical-rfi-removal,pieflag-middelberg-2006,offringa-2010-post-correlation-rfi-classification,prasad-flagcal-2012,serpent-peck-2013,nn-rfi-detection-2020,sun-2022}). This process is referred to as data flagging, and is also our method of choice for dealing with RFI in Apertif data in this work. Our detection methodology builds upon the RFI detection pipelines for the Low-Frequency Array (LOFAR; \citealt{lofar-2013,LOFAR-RFI-pipeline}) and the Murchison Widefield Array (MWA; \citealt{mwa-2013,offringa-2015-mwa-rfi}). Those pipelines integrate an \textsc{aoflagger} flagging strategy, which combines filtering, sumthresholding, morphological operations and heuristics. Details of the \textsc{aoflagger} approach will be discussed in \S\ref{sec:aoflagger-method}.

\begin{figure*}
  \begin{center}
	\includegraphics[width=13cm]{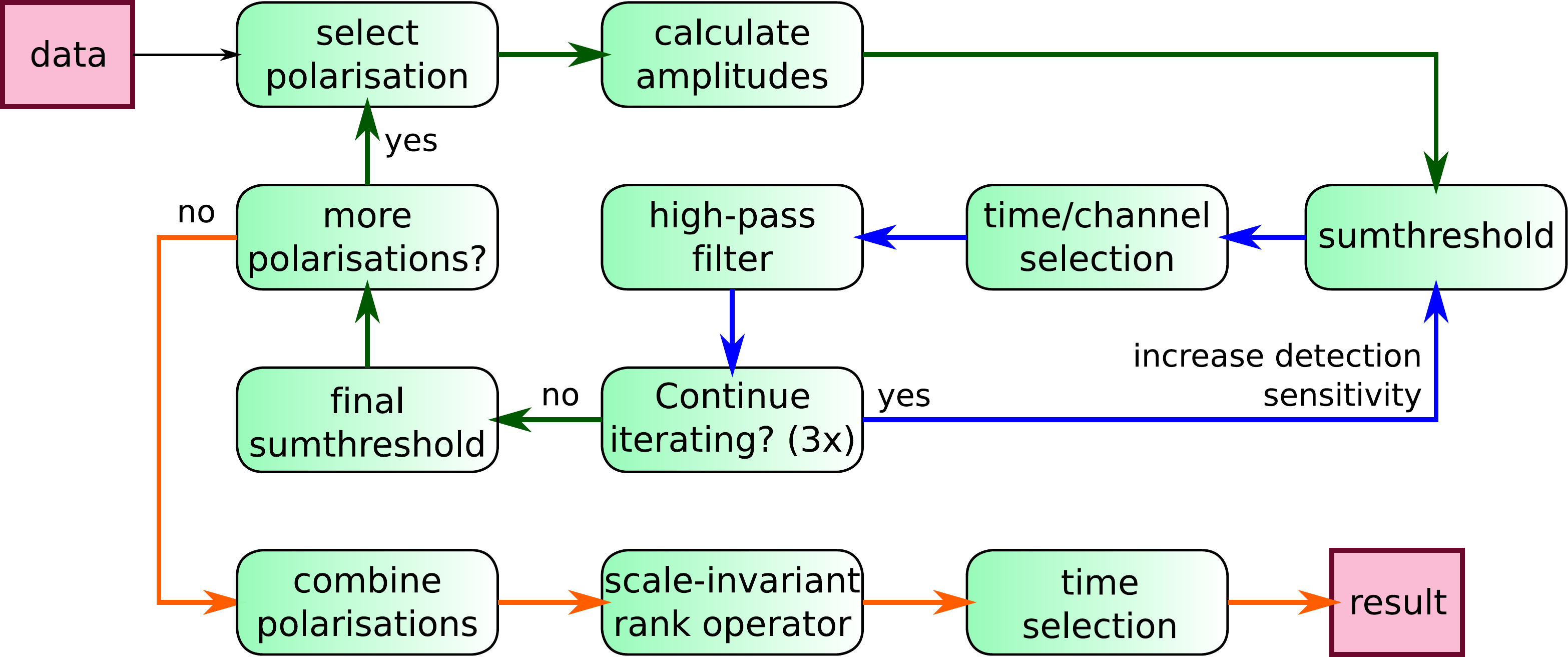}%
	\caption{The default \textsc{aoflagger} strategy for RFI-detection (before modifications for Apertif). These steps are independently performed on smaller subsets of the data. The input data of one independent run through these steps typically consists of approximately an hour of correlations from a single pair of antennas and a single beam, with the full bandwidth and all four linearly polarized cross-correlations present.}
	\label{fig:aoflagger-diagram}
  \end{center}
\end{figure*}

Apertif supports a transient (beam-formed) mode and an imaging mode. The RFI detection approach for these two modes are fundamentally different. In this work we aim at RFI detection in imaging mode, i.e., after having correlated and integrated the voltages from all the antennas. See \citet{sclocco-2019} for an approach to mitigate RFI in beam-forming mode. Our approach is part of a fully automated Apertif imaging pipeline called \textsc{apercal} \citep{adebahr-apercal}.

A multi-beam receiver makes it possible to perform spatial filtering techniques to suppress interference \citep{spatial-filtering-parkes-multibeam,spatial-filtering-parkes-multibeam-for-pulses,rfi-spatial-processing-hellbourg-2014}. This requires fast dedicated computing hardware that processes the raw signals from all the beams, which for Apertif is not available. Spatial filtering is also mainly used to filter out a limited number of known transmitters, which for Apertif is likely not sufficient by itself, although it might save some part of the bandwidth.

Another approach to detect interference is by using the spectral kurtosis statistic \citep{hardware-implementation-spectral-kurtosis,taylor-2019,purver-2021}. This has shown results that are competitive with amplitude-based detection. However, this requires a specialized correlator and a doubling of the data volume to be able to calculate the kurtosis.

Recently, machine learning has been used to address the issue of RFI detection \citep{harrison-2019,nn-rfi-detection-2020,xiao-2022,sun-2022}. \citet{nn-rfi-detection-2020} argue that convolutional neural networks can achieve an accuracy that is higher than that of their {\sc sumthreshold} implementation. For this comparison, the authors use their own customized implementation of the {\sc sumthreshold} method, whereas in platforms such as \textsc{aoflagger} the method is typically applied iteratively and combined with filters \citep{offringa-2010-post-correlation-rfi-classification,LOFAR-RFI-pipeline} and morphological operators \citep{offringa-2012-scale-invariant-rank-operator,vdgronde-siroperator-2016} to enhance the accuracy. With these additions, it has been shown that pipelines such as \textsc{aoflagger} typically detect all interference that astronomers would manually flag. In this work, we will showcase what can be achieved with traditional methods --- including their computational requirements --- thereby providing an updated base-line to compare against.

In this paper, we introduce a flagging strategy for Apertif data using the AOFlagger framework, and demonstrate our designed strategy on Apertif data. In \S\ref{sec:method}, we will start by introducing the AOFlagger steps used to construct the Apertif approach, and introduce several new operations that are integrated into the Apertif flagging strategy. In \S\ref{sec:results}, results of applying this strategy are presented, including long-term statistics and the computational requirements. Finally, in \S\ref{sec:discussion-conclusions} we discuss the results and draw conclusions.

%--------------------------------------------------------------------
\section{Method} \label{sec:method}
For this work, we have designed an interference detection approach for Apertif based on the existing \textsc{aoflagger} approach and integrated this into the \textsc{apercal} pipeline. \textsc{apercal} is an automated processing pipeline for Apertif imaging observations \citep{adebahr-apercal}, consisting of common steps such as data formatting, interference detection, calibration and imaging. Interference detection is one of the first steps during data reduction and is fundamental for achieving a good and persistent calibration and image quality and later steps of the processing.

To improve the detection quality, several modifications to \textsc{aoflagger} are required. This consists of extensions of existing algorithms and optimizing parameters for \textsc{apertif}, which we will discuss in this section. We will start with an overview of the detection approach.

\subsection{Overview} \label{sec:aoflagger-method}
Fig.~\ref{fig:aoflagger-diagram} shows an overview of the steps that the default \textsc{aoflagger} strategy performs. The \textsc{aoflagger} approach to RFI detection in a subset can be summarized as i) estimation and subtraction of the sky signal by applying a Gaussian high-pass filter in time-frequency space (see \S\ref{sec:high-pass-filter}); and ii) detection of excessive values, with increased sensitivity towards spectral-lines and broadband features. The detection is performed with the \textsc{sumthreshold} algorithm \citep{offringa-2010-post-correlation-rfi-classification}. Steps i) and ii) are typically iterated three times with increased sensitivity to make sure that the final sky signal estimate is minimally biased by interference. As a final step, the flags from different polarizations are combined and are extended in time and frequency, using the scale-invariant rank (SIR) operator \citep{offringa-2012-scale-invariant-rank-operator,vdgronde-siroperator-2016}. This latter step improves detection of interference that tapers off below the noise floor and fills gaps in the flag mask when a persistent transmitter is not fully detected.

With \textsc{aoflagger}, detection of interference is performed independently on subsets of the data, and the pipeline of Fig.~\ref{fig:aoflagger-diagram} runs independently for each subset. For Apertif, such a subset was chosen to contain the data from all four linearly polarized correlations (XX, XY, YX, YY), the full bandwidth (300~MHz), an interval of typically half an hour for a single beam and a single correlated baseline. Hence, the detection of interference for different beams, baselines and time intervals is independently performed, even though these are part of the same observation. The motivation for flagging these subsets independently is two fold:
\begin{itemize}
 \item It improves performance: it allows parallel and distributed detection of subsets. The independent flagging of beams and time intervals matches with the format of the data. Despite this, data access is still not ideal, because the data for one baseline is stored dispersed over the time direction.
 \item Combined detection does not significantly improve detection: the added value of detection on combined subsets of data is small, i.e., one subset contains little information about the RFI in another subset. This is because the impact of RFI can vary greatly between different beams and different baselines. Furthermore, it rarely occurs that RFI which affects image quality is not detectable in half an hour of data, but is detectable when multiple half hour intervals are combined.
\end{itemize}

Performing detection on integrated baselines has, in some cases, been shown to make faint RFI detectable \citep{offringa-2015-mwa-rfi,wilensky-2019}. Early tests with Apertif data, however, indicated that there is no gain in combining baselines. We have also performed tests that flag after integrating over multiple beams, but again found no improvement in doing so. These tests were not exhaustive and it could be that combined detection on baselines or beams could still improve the accuracy somewhat.

\textsc{aoflagger} aims to take out RFI that requires raw, high-resolution data flagging. Because of the high resolution of processed data, the computational performance of detection is critical. It is important to perform high-resolution flagging early, because it results in the highest accuracy and the impact of flagging is reduced compared to low-resolution flagging \citep{offringa-lofar-environment-2013}. On the other hand, some phenomena cause the loss of large time intervals or frequency ranges. Common instrumental causes are correlator failures, temporary local RFI or strong broadband transmitters. Detection of such issues does not require the high-resolution data, and it is therefore less critical to detect such issues in the first \textsc{aoflagger} detection run. Such issues can be found in post-processing of lower-resolution data for which the performance is less critical.

\subsection{Invalid data}
There are several instrumental issues that may result in data with invalid values that interrupt the data in time or frequency. A few examples of such issues are correlator malfunctions, dish shadowing, incorrectly set sub-band gains, network failures (between stations and the correlator) or data corruption. Such instrumental issues result in visibilities that may have non-physical values for certain times, frequencies, feeds or antennas, or could lead to visibilities with a not-a-number (NaN) value. We will refer to such data as invalid data.

In most cases, invalid data can be detected and flagged early in the processing. For example, shadowing can be determined from the target direction and the layout of the array, and missing sub-band data caused by network congestion can be detected by the correlator. In this paper, we consider the detection of such issues outside the context of interference detection. It does, however, make it necessary for the detector to continue to work in the presence of (pre-detected) invalid data, which may affect only specific times, frequencies or some other selection of data.

Making the \textsc{aoflagger} algorithm aware of invalid data is one of the changes that was required for Apertif. The \textsc{aoflagger} algorithm was originally designed to work on raw high-resolution single-subband LOFAR data. It rarely happens that such a span of data is partially invalid, and initially \textsc{aoflagger} algorithms therefore do not take invalid data into account. In the case of Apertif, the full bandwidth is offered to \textsc{aoflagger}, and the loss or corruption of a single subband causes therefore gaps in the bandwidth. Being a different instrument, Apertif is also affected by different issues that may not affect LOFAR, such as shadowing. For these reasons, we have extended the \textsc{aoflagger} algorithm to take invalid data into account. This requires changes to the \textsc{sumthreshold} and \textsc{sir-operator} steps of the algorithm, which we will discuss in the next two sections.

\subsection{Extension of the \textsc{sumthreshold} algorithm}
\begin{figure*}
  \begin{center}
	\includegraphics[width=10cm]{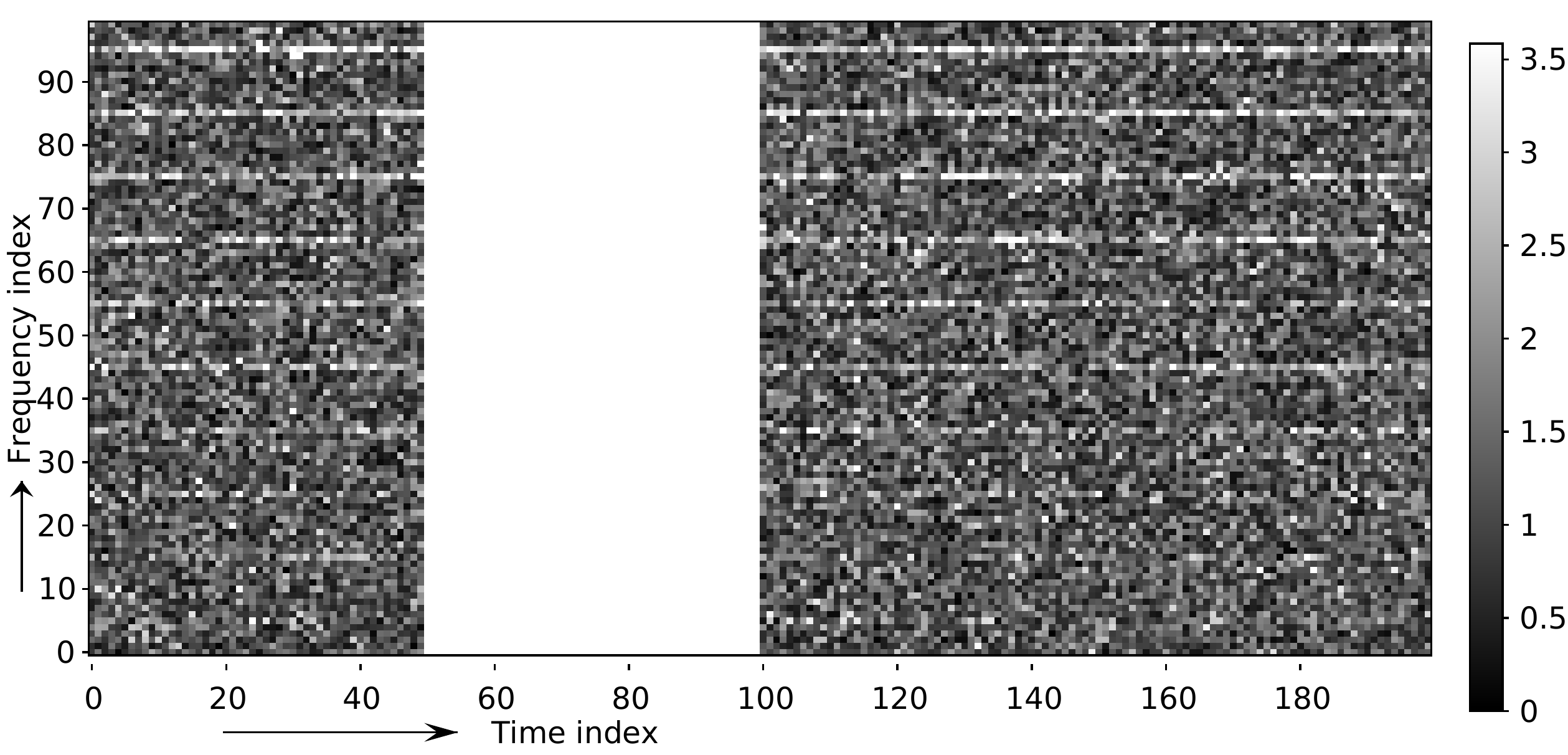}\\%
	\includegraphics[width=5cm]{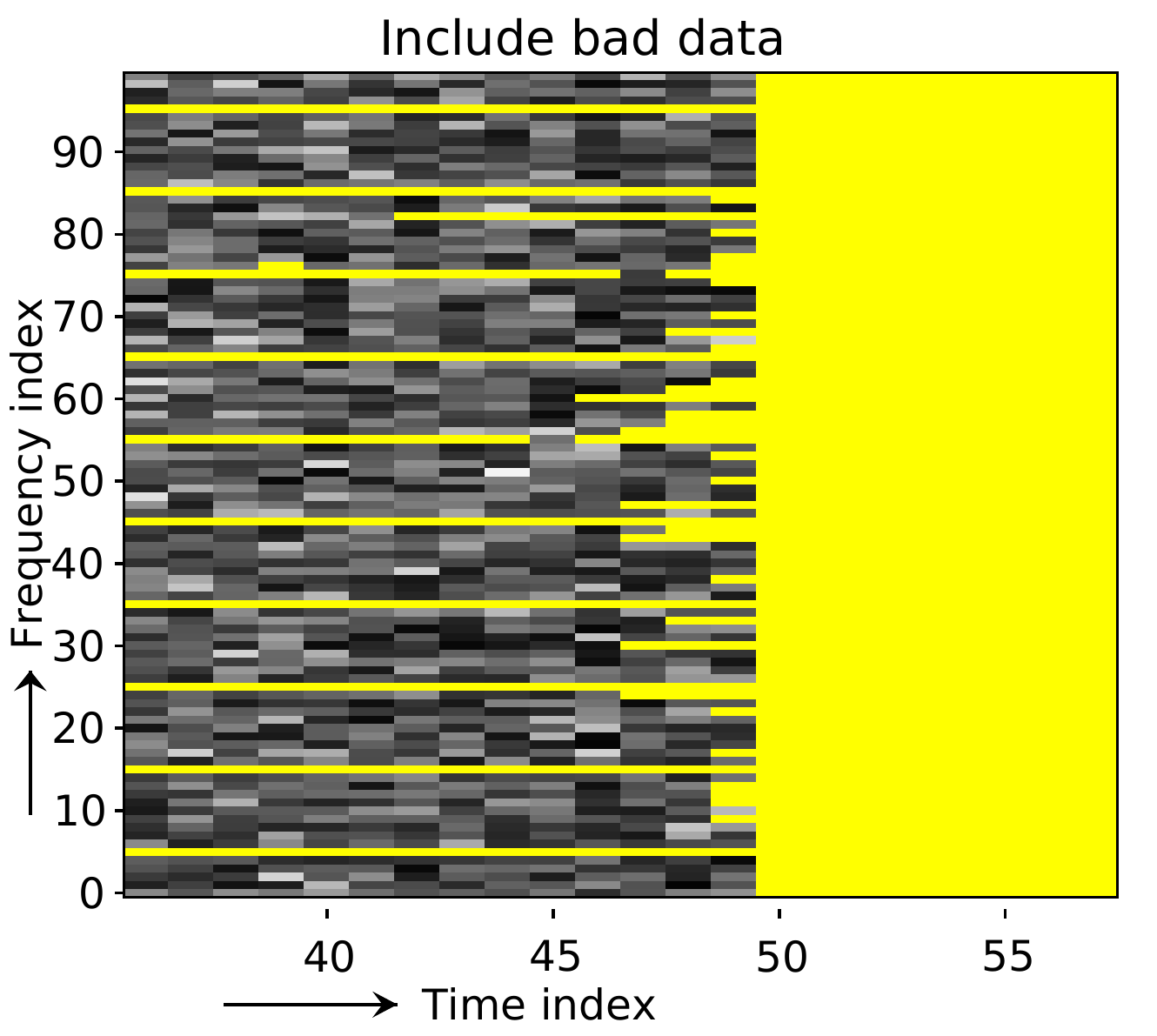}%
	\includegraphics[width=5cm]{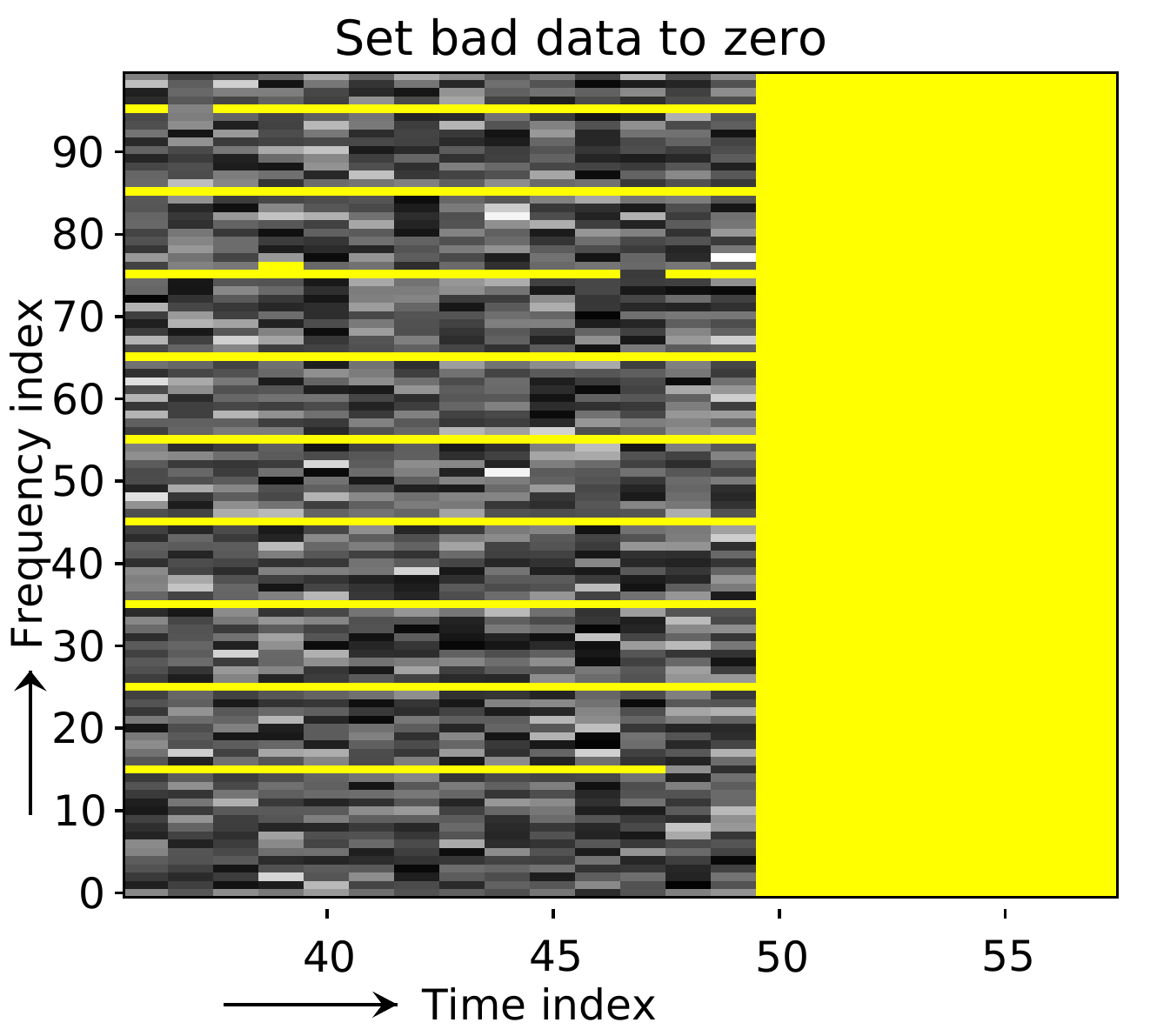}%
	\includegraphics[width=5cm]{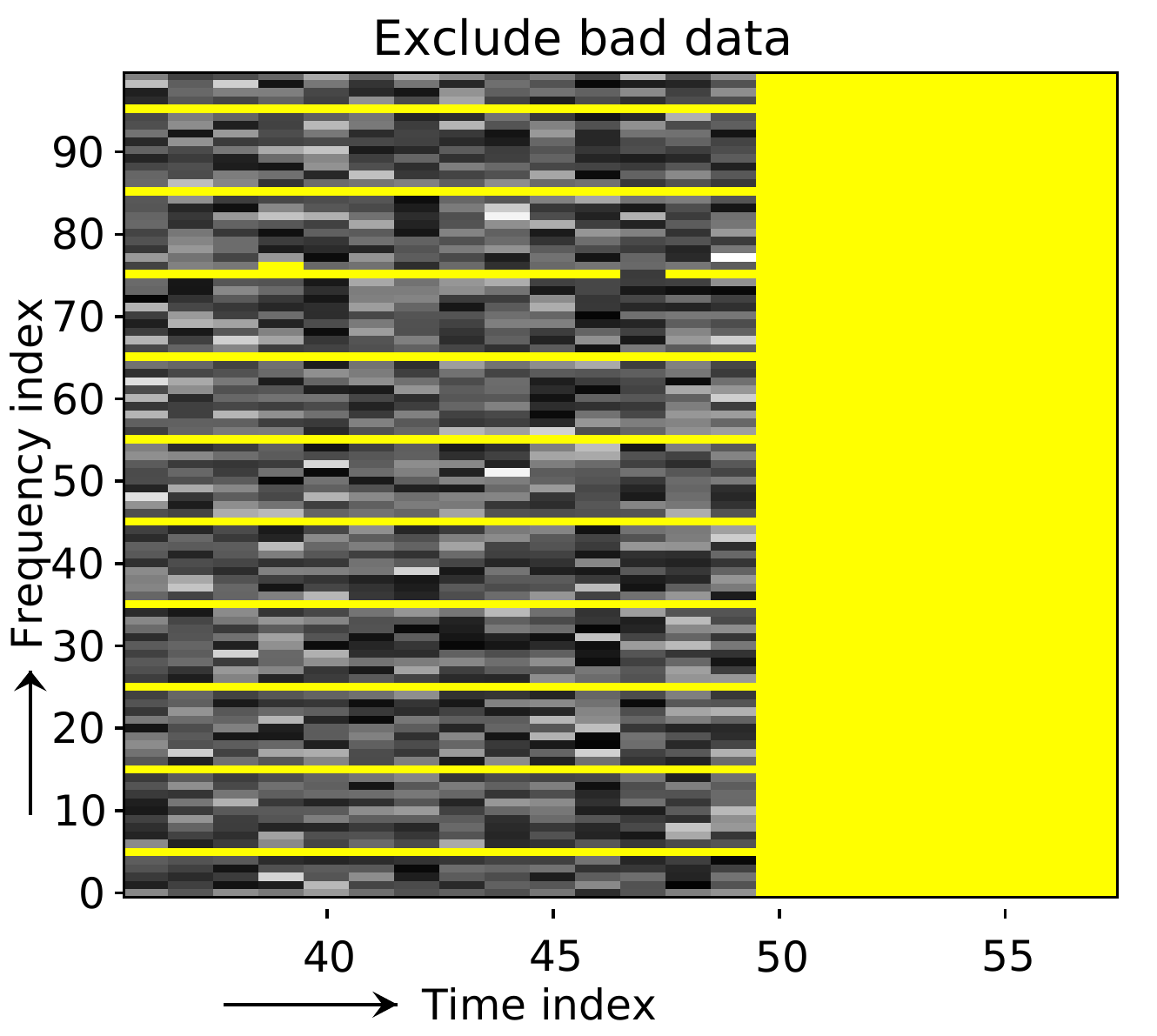}%
	\caption{Three methods of handling invalid data in the \textsc{sumthreshold} step. The top image shows the simulated input data, which consists of Gaussian complex noise, spectral line RFI every 10 channels that increases in strength in frequency direction, and a block of invalid data (time indices 50--100), simulating e.g. a temporary correlator failure. The bottom images show a zoom in on the left edge of the invalid data. Flagged data is marked in yellow. Bottom-left: normal \textsc{sumthreshold} without using knowledge of the invalid data; bottom-centre: invalid samples are set to zero before \textsc{sumthreshold}; bottom-right: invalid samples are removed before \textsc{sumthreshold}.}
	\label{fig:sumthreshold}
  \end{center}
\end{figure*}
The \textsc{sumthreshold} algorithm is a combinatorial thresholding method that detects line-like structures in the time-frequency data \citep{offringa-2010-post-correlation-rfi-classification}. This method is effective for the detection of RFI, because most RFI raises the amplitude of consecutive time or frequency samples. The method iteratively thresholds the average over an increasing number of neighbouring samples with a decreasing threshold. With $i$ the zero-indexed iteration number, $M_i$ the number of samples, $\chi_i$ the threshold and $\rho$ a constant normally chosen to be 1.5,
\begin{eqnarray}
 M_i    &=& 2^i\\
 \chi_i &=& \chi_0 \, \rho^{-\log_2 M_i }.
\end{eqnarray}
$\chi_0$ is a user parameter that controls the total sensitivity of the method. The various default \textsc{aoflagger} algorithms use values of $\chi_0=$ 6\ldots8.5 $\sigma$. The mode of the noise $\sigma$ is determined from the data that is (at that point in the detection) determined to be RFI free, and is estimated by calculating the truncated mode of the RFI free data, skipping 20\% of the outlier values (the 10\% minimum and maximum values), thereby assuming that the inner 80\% follow a Rayleigh distribution. Assuming that the contribution of the noise is Gaussian distributed in the real and imaginary components of the visibilities, this results in a stable estimate of its standard deviation \citep{fridman-variance-estimates-2008}.

A single iteration consists of thresholding all sequences of size $M_i$ in both the time and the frequency direction (unless $M_i=1$), possibly with different thresholds for the two dimensions, to separately control the sensitivity towards spectral line RFI and transient broadband RFI. Typically, a total of 9 of these iterations are performed, giving a maximum size of $M_8=256$. A sample that is flagged in an earlier iteration or direction, is (temporarily) replaced by the mean of the non-flagged samples in the sequence. The following description demonstrates the first three iterations, using $\chi_0=6$ and $\rho=1.5$:
\begin{enumerate}
 \item Flag samples with an absolute value $\ge 6 \sigma$.
 \item \begin{enumerate}
 \item Flag every sequence of 2 consecutive samples in time with an absolute average $\ge 4 \sigma$ (because $\chi_2=6\sigma\times1.5^{-\log_2 (2^2)} = 4\sigma$).
 \item Flag every sequence of 2 consecutive samples in frequency with an absolute average $\ge 4 \sigma$.
\end{enumerate}
 \item Repeat 2.(a) and (b) with 4 samples and a threshold of $\chi_4=6\sigma \times 1.5^{-\log_2 (2^3)} = 2\frac{2}{3}\sigma$.
\end{enumerate}
Subsequent iterations will threshold sequences of 8, 16, 32, $\ldots$ samples with an average above $\chi_8 \approx 1.8 \sigma$, $\chi_{16} \approx 1.2 \sigma$, etc.

In the form described by \citet{offringa-2010-post-correlation-rfi-classification}, pre-existing classification of invalid data is not taken into account in the \textsc{sumthreshold} method. An example of such a case is shown in Fig.~\ref{fig:sumthreshold}, which considers a simulated observation with 200 timesteps and 100 channels. The observation contains spectral-line interference that affects one channel out of every ten channels and increases power at higher frequencies. Timesteps 50---100 are known to be invalid data, and are set to high values by raising them with 10 times the standard deviation.

The second row of Fig.~\ref{fig:sumthreshold} zooms in on time indices 30-60. The first image of the second row shows the result of a basic application of \textsc{sumthreshold}. For this result, the knowledge that some data was invalid is not used. As a result, the invalid data is considered to be RFI, and samples before and after the block of invalid data are flagged with an increased sensitivity. As a result, the false-positive rate is clearly increased.

A simple approach to mitigate this is to consider invalid values to be zero when applying the \textsc{sumthreshold} method. This results in the plot shown in the middle of the second row of Fig.~\ref{fig:sumthreshold}. This result does not show increased false positives because of the invalid data. With this approach, information about flagged samples on either side (before/after) of the missing data does not (significantly) aid detection, because the invalid data is considered to be zero, and this lowers the average absolute sum in the iterations of the \textsc{sumthreshold} method that consider longer consecutive ranges. This results in a higher false-negative rate than would theoretically be possible if the information on both sides of the invalid data would have been used together. In particular, the faintest interfering line at channel index 5 is no longer detected.

While the loss in accuracy is minimal, there is a simple method to aid the detection of interference on one side of the block of invalid data with information from the other block: by completely skipping data in the summed direction (time or frequency). In other words, samples that are directly before and after a block of invalid data are treated as if they are consecutive. The result of this is shown in the third column of Fig.~\ref{fig:sumthreshold}, which indeed shows a lower false-negative rate. In particular, the faintest spectral line at channel 5 is now fully detected.

When comparing these two approaches to deal with invalid data, the approach to exclude the invalid data leads to a small increase in false-positive detections when the RFI is not consistently present in time or frequency, i.e. when it is present on one side of the invalid data block and not present on the other side. This should be weighted against the increased sensitivity when the RFI is consistently present. The optimal choice therefore depends on the behaviour of the RFI. Because persistent RFI is common, and because it is more important to avoid false negatives in persistent RFI (which might negatively affect later processing steps) over avoiding false negatives in transient RFI (which would lead to a small loss of data), we use the method of excluding invalid data in our Apertif strategy.

We have implemented this in two ways: i) stack all valid data into a temporary storage, run the normal \textsc{sumthreshold} algorithm on these data and reverse the stacking operation on the resulting mask; and ii) skip over the invalid data inside the \textsc{sumthreshold} method. We have timed these two implementations on simulated complex Gaussian data with 10,000 timesteps $\times$ 256 channels. Each run is repeated 100 times. The first implementation runs about $2.5\times$ faster (0.18~s per data set) compared to the second implementation (0.45~s per data set). The first method is still $6\times$ slower compared to the regular algorithm (which takes 0.03~s per data set). This can be explained by the extra copying of data that is required in each iteration (both for the time and for the frequency direction).
% Vertically stacking: 13.783 sec / 10
% Consecutive: 18.021 / 10
% Reference: 16.772 / 10

\subsection{Extension of the \textsc{scale-invariant rank operator}}
The SIR-operator is a morphological operation that is used in \textsc{aoflagger} to extend the detected RFI mask in the time and frequency direction. It is an effective step to follow threshold-based methods to detect faint RFI based on the morphology of detected flags \citep{offringa-2012-scale-invariant-rank-operator, vdgronde-siroperator-2016}. It is scale invariant, which implies that the fractional increase in flags in one dimension is constant, i.e., independent of the scale of that feature in that dimension. 

The SIR-operator is essentially a one-dimensional operator that can be applied to a sequence of flag values. To apply it to radio interferometric data, \citet{offringa-2012-scale-invariant-rank-operator} apply the operator in both the time and frequency dimensions: in time it is separately applied to all the channels, and in frequency it is applied separately to all timesteps. The union of these to steps is taken as the result.

Assume that $X$ is a single sequence of flag values, such that $X[i]$ holds a Boolean value that represents the state of the flag. The output $\rho(X)$ of the SIR-operator applied to $X$, is defined as the union of all subsequences of the input $X$, for which
\begin{equation} \label{eq:sir-operator-eq}
\#^{i:j}_{\mathcal{F}} \ge (1 - \eta) \left( j - i \right).
\end{equation}
Here, $\#_{\mathcal{F}}^{i:j}$ is brief for $\#_{\mathcal{F}}(X[i : j])$, which is the count-operator that returns the number of flagged samples in a sequence. $X[i : j]$ is the subsequence of samples consisting of all elements $X[k]$ for $i \le k < j$ and $\eta \in [0 \ldots 1]$ is a tunable parameter that sets the aggressiveness of the operator.

\begin{figure*}
  \begin{center}
	\includegraphics[width=8cm]{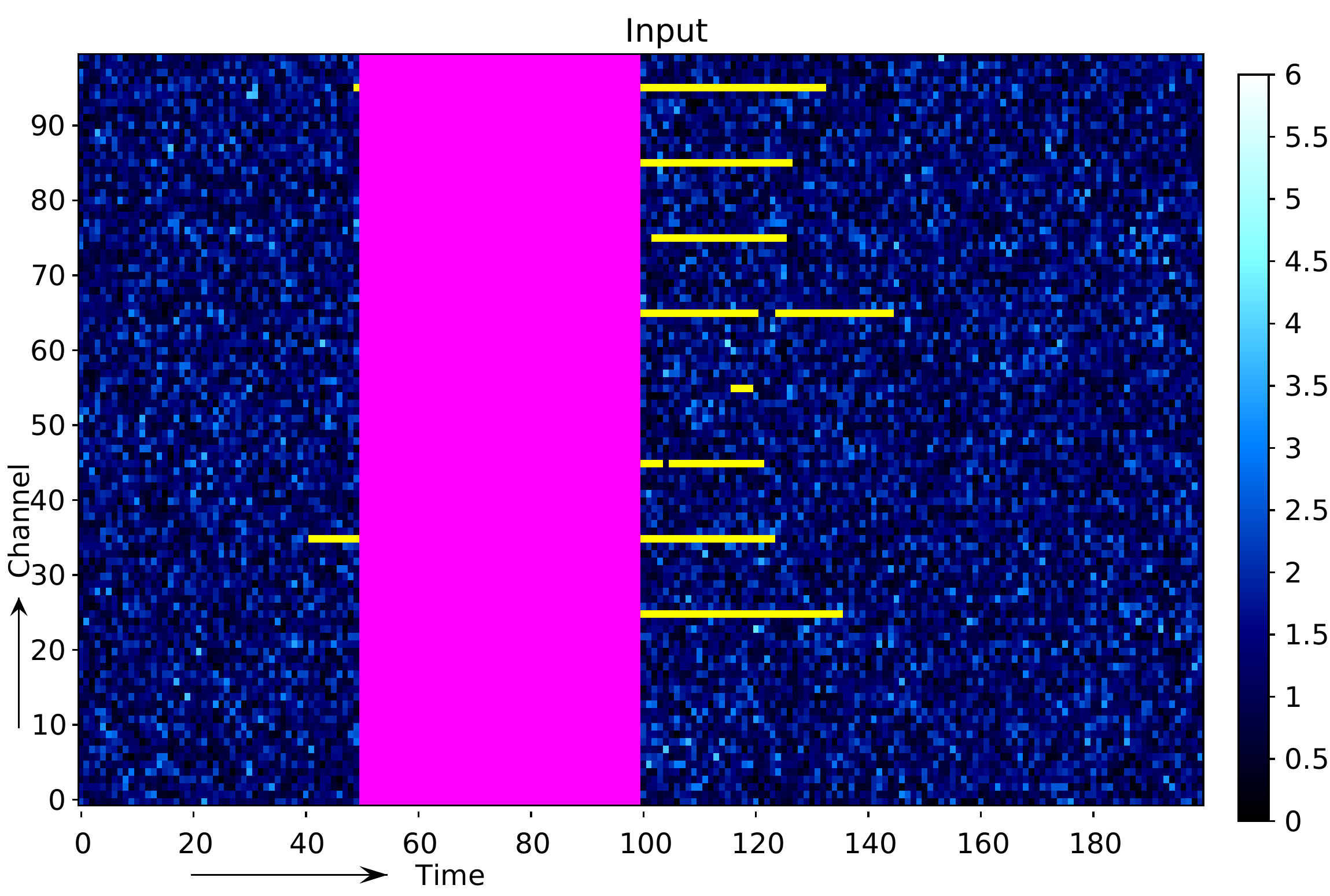}\\%
	\includegraphics[width=8cm]{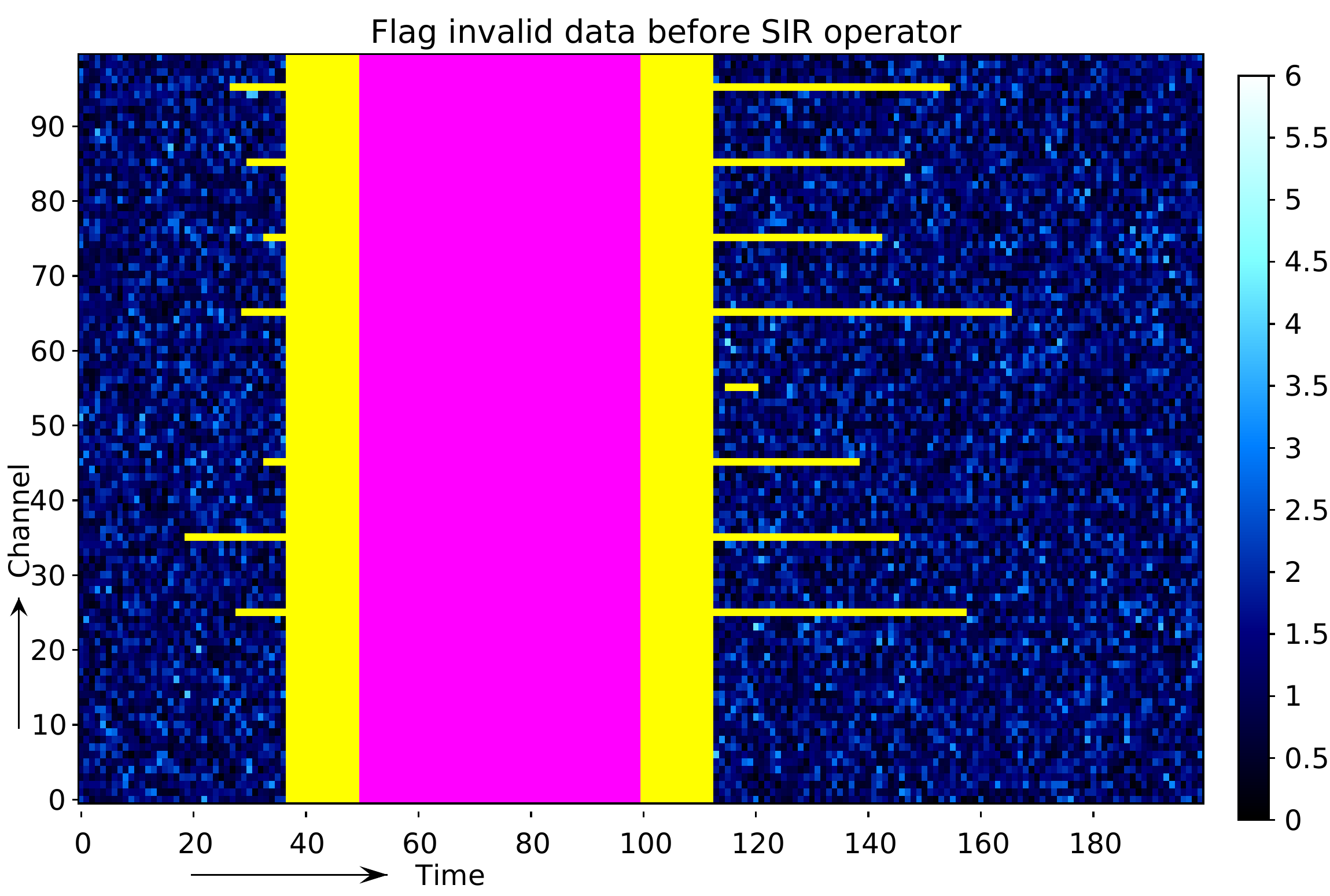}%
	\includegraphics[width=8cm]{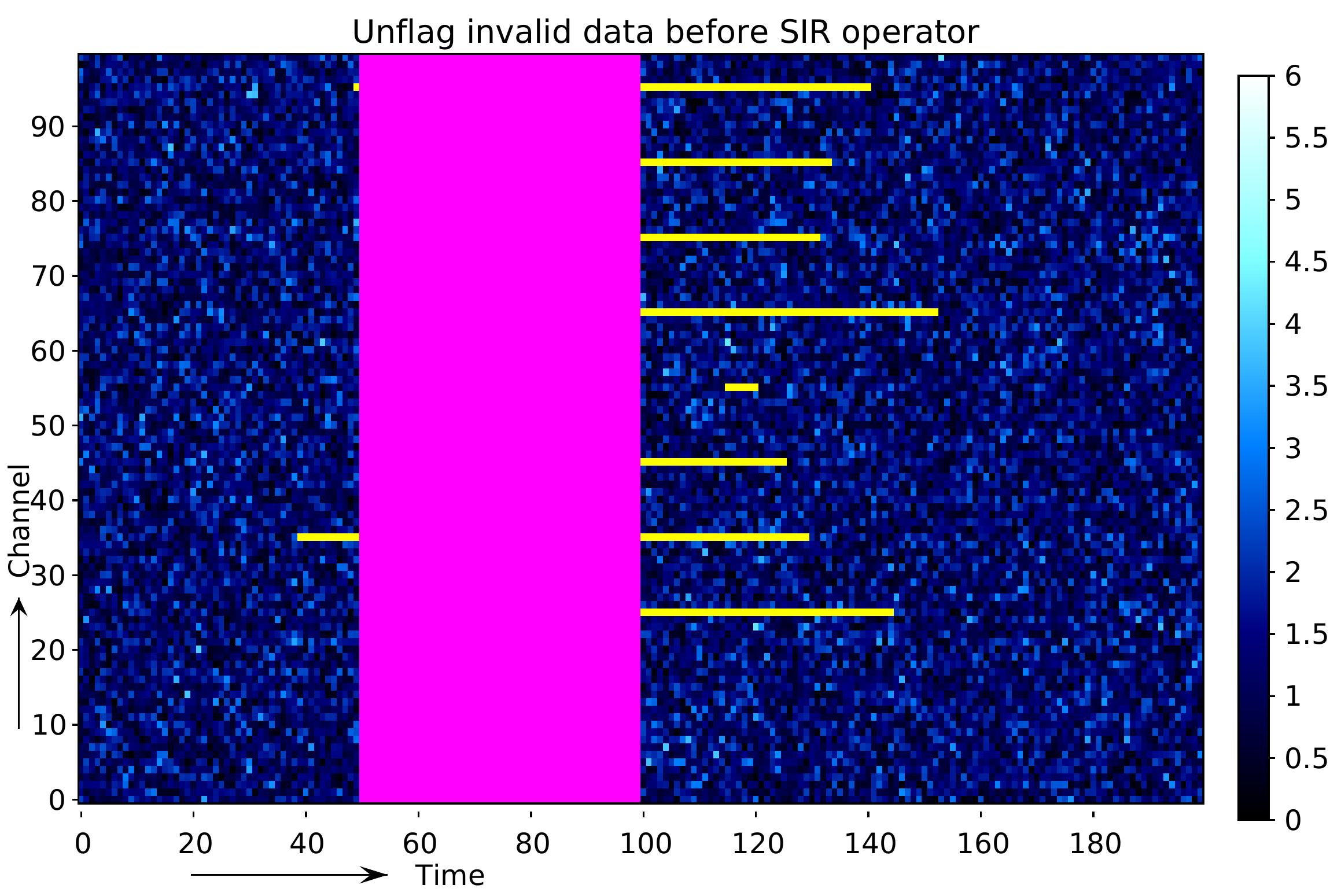}\\%
	\includegraphics[width=8cm]{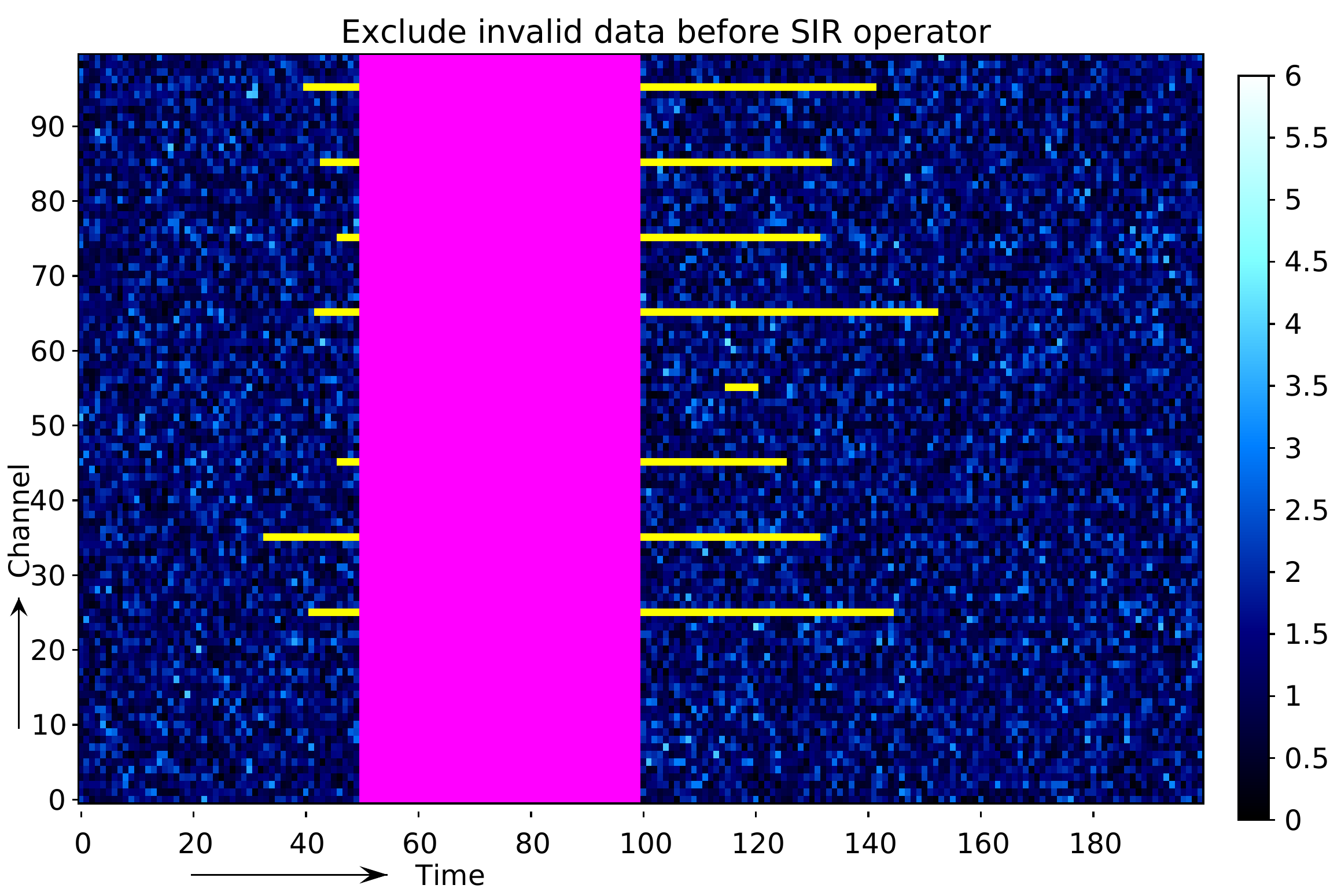}%
	\includegraphics[width=8cm]{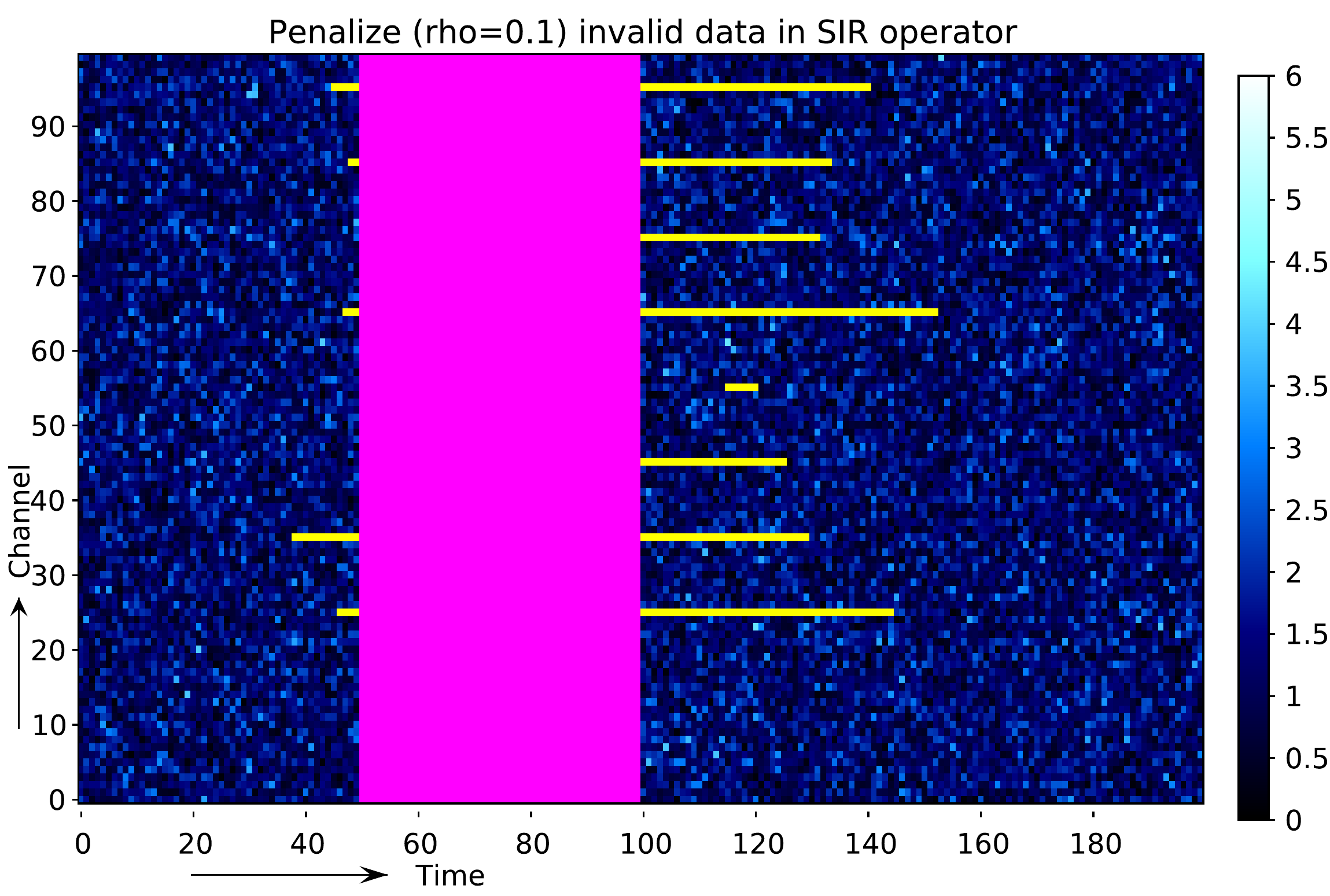}%
	\caption{Different ways of handling invalid data in the \textsc{sir-operator} step on a simulated data set with a Gaussian burst of interference in a few channels. Purple marks invalid data, yellow is detected as interference. The SIR-operator operates on the flag mask, hence the visibility values are not used. Top: input data. Centre-left: Invalid data is counted as flagged data. Centre-right: Invalid data is counted as unflagged data. Bottom-left: Invalid data is removed before applying the SIR-operator. Bottom-right: Invalid data is penalized with $\rho=0.1$.}
	\label{fig:sir-operator-gaus}
  \end{center}
\end{figure*}

Eq.~\eqref{eq:sir-operator-eq} implies that a sequence of flags caused by invalid data is extended on both sides by a ratio of $\eta$. An example of this is given in the centre-left panel of Fig.~\ref{fig:sir-operator-gaus}. This behaviour is undesirable because, unlike most RFI signals, invalid data typically has a sharp boundary and should be flagged like that. The extension of invalid data causes a high number of false positives.

A simple solution is to count invalid data as unflagged data in the SIR operator. This implies that Eq.~\eqref{eq:sir-operator-eq} is modified so that the count operator only counts the number of flags corresponding to valid data:
\begin{equation} \label{eq:sir-operator-unflag-invalid}
\#^{i:j}_\mathcal{FV} \ge (1 - \eta) \left( j - i \right),
\end{equation}
where $\#_\mathcal{FV}$ is the number of valid samples that are flagged in the interval $X[i:j]$ (as opposed to $\#_\mathcal{F}$, which counts flagged values that can both be valid or invalid). Because the right side is unchanged and the left side remains equal or is decreased compared to Eq.~\eqref{eq:sir-operator-eq}, this modification always flags an equal or fewer amount of samples. An application of this approach is demonstrated in the centre-right panel of Fig.~\ref{fig:sir-operator-gaus}. This approach remedies the extending of flags around invalid data.

The downside of the approach of Eq.~\eqref{eq:sir-operator-invalid-skipped} is that a continuous transmitter is assumed not to be present in the invalid data range, causing flags on either side to have a decreased probability of getting flagged. For example, in case a correlator fails for a minute during which a transmitter remains present in one channel with decreasing power, the transmitter is less likely to be flagged after the correlator failure. To address this, we further modify Eq.~\eqref{eq:sir-operator-eq} to:
\begin{equation}\label{eq:sir-operator-invalid-skipped}
 \#_\mathcal{F}^{i:j} \ge (1 - \eta) \, \#_\mathcal{V}^{i:j},
\end{equation}
where $\#_\mathcal{V}^{i:j}$ is the number of valid (flagged or unflagged) samples in interval $X[i:j]$. This approach is effectively the same as removing the invalid samples from the sequence before applying Eq.~\eqref{eq:sir-operator-eq}. Therefore, a transmitter that gets interrupted by invalid data receives a higher probability to get flagged. An example of this approach is given in the bottom-left panel of Fig.~\ref{fig:sir-operator-gaus}. Invalid samples are skipped in this approach, and flagged samples on one side of a sequence of invalid samples may increase the probability of samples on the other side of the sequence, irregardless of the size of the invalid sample sequence.

The approach of Eq.~\eqref{eq:sir-operator-invalid-skipped} can overstep its goal of using information from before and after a sequence of invalid data, in particular in the case of very long sequences of invalid samples. For example, when considering a transmitter that is active for one minute before the receiving antenna is shadowed for 6 hours (causing invalid data), it is undesirable that samples after shadowing receive higher detection probability because of what happened 6 hours ago. A final modification to the SIR operator we consider is therefore to introduce a penalty parameter $\rho$ that can balance between Eqs.~\eqref{eq:sir-operator-unflag-invalid} and \eqref{eq:sir-operator-invalid-skipped}:
\begin{equation}\label{eq:sir-operator-penalty}
 \#_\mathcal{F}^{i:j} \ge (1 - \eta) \left((j-i)\rho + \, \#_\mathcal{V}^{i:j} (1-\rho) \right).
\end{equation}
With $\rho=0$, invalid samples are skipped, making the method equal to Eq.~\eqref{eq:sir-operator-invalid-skipped} and with $\rho=1$, invalid samples are counted as unflagged samples, making the method equal to Eq.~\eqref{eq:sir-operator-unflag-invalid}. A value of $\rho=0.2$ implies that five invalid samples count as one unflagged sample, thereby lowering the probability of flagging through a block of invalid data, but still transferring some of the flag information from before to after the invalid data and vice versa. This method is demonstrated with a setting of $\rho=0.1$ in the bottom-right panel of Fig.~\ref{fig:sir-operator-gaus}. 

Considering the results of all approaches in Fig.~\ref{fig:sir-operator-gaus}, it is clearly undesirable to generally extend invalid data using the traditional SIR-operator defined in Eq.~\ref{eq:sir-operator-eq}. Any of the three different variations of the algorithm (Eqs.~\ref{eq:sir-operator-unflag-invalid}, \ref{eq:sir-operator-invalid-skipped} and \ref{eq:sir-operator-penalty}), which can be described by choosing different $\rho$-values in Eq.~\eqref{eq:sir-operator-penalty}, solve this issue. The different values of $\rho$ do not cause significant changes. We have tested values of $\rho$ on a few observations, some with artificially added invalid data, and visually compared the flagging results. Based on these results and the arguments given earlier about finding a balance between Eqs.~\eqref{eq:sir-operator-unflag-invalid} and \eqref{eq:sir-operator-invalid-skipped}, we use $\rho=0.1$.

Introducing the parameter for invalid-data weighting $\rho$ has no significant effect on the speed of the algorithm. The original algorithm can be implemented with a computational complexity of $\mathcal{O}(N)$ \citep{offringa-2012-scale-invariant-rank-operator}, and the same holds for the algorithm that includes the invalid-data penalty parameter.

\begin{figure*}
  \begin{center}
	\includegraphics[width=8cm]{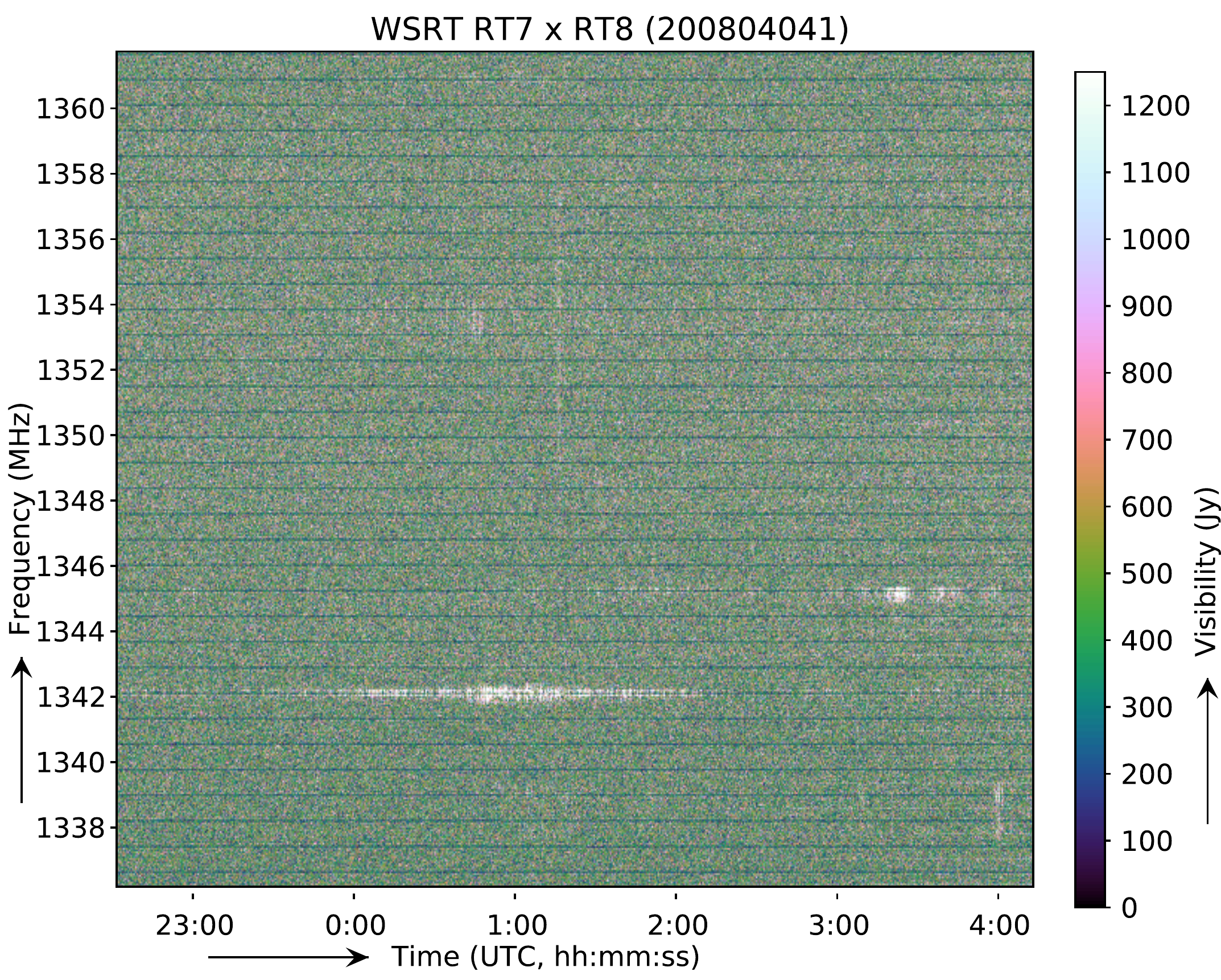}%
	\includegraphics[width=8cm]{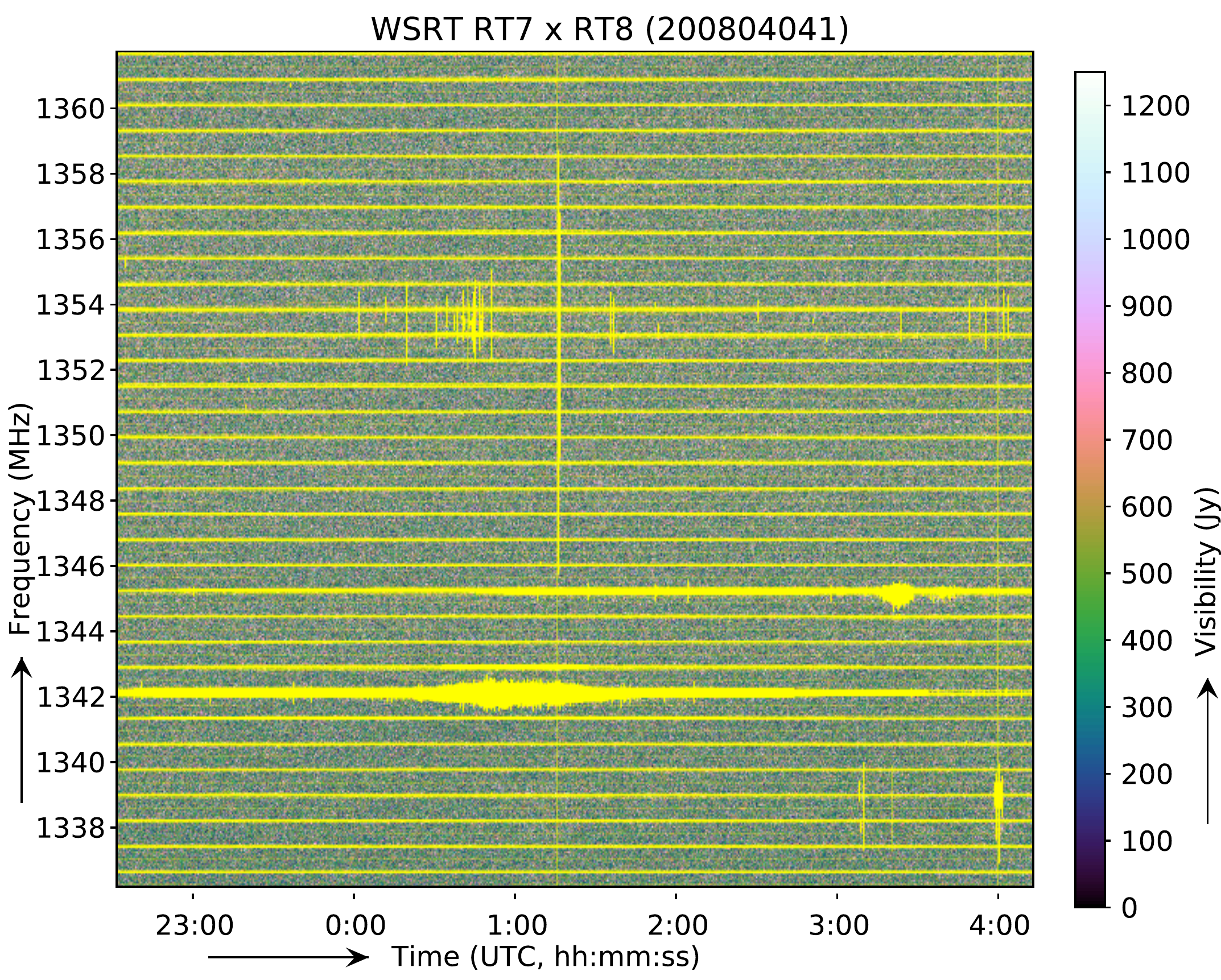}\\%
	\includegraphics[width=8cm]{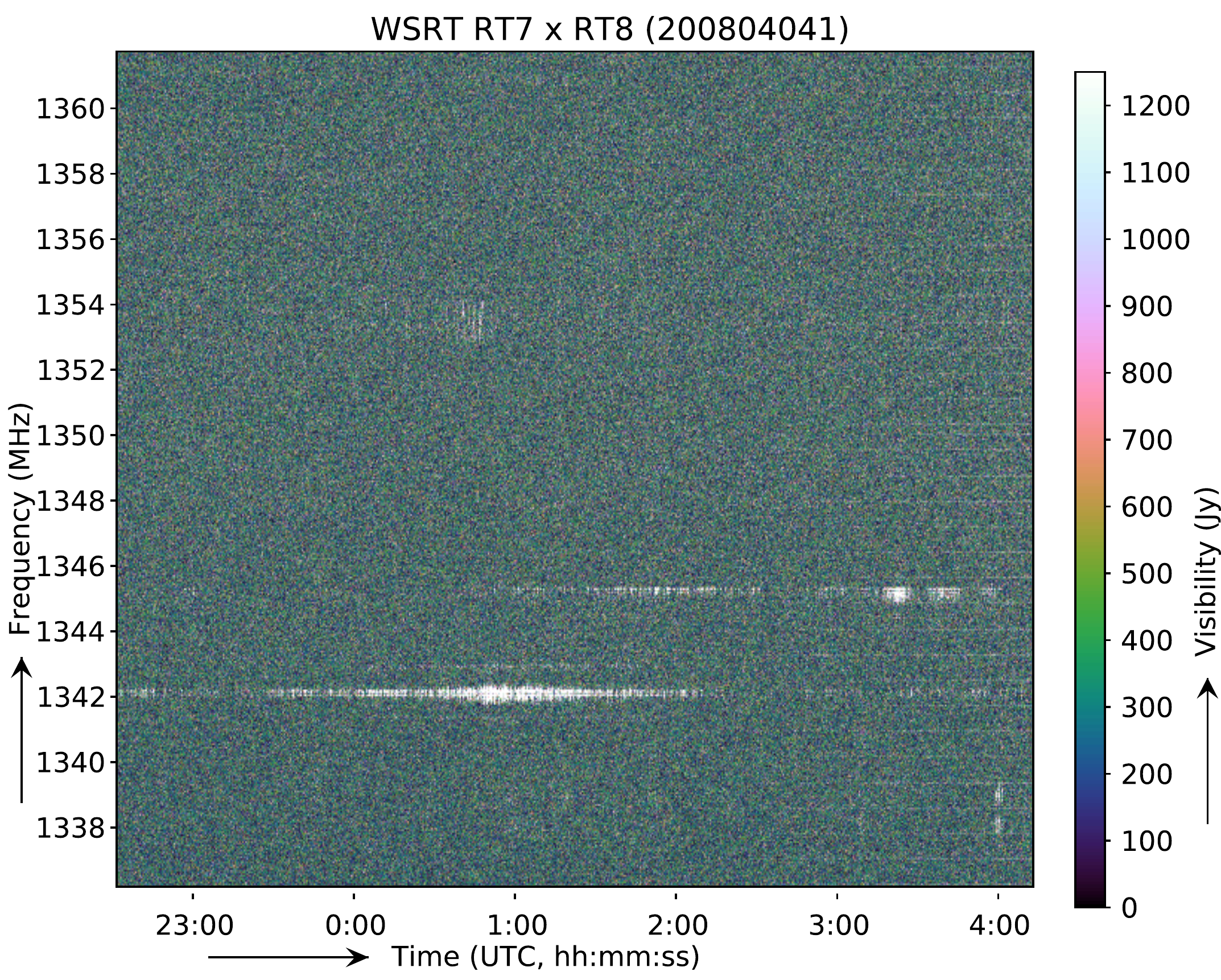}%
	\includegraphics[width=8cm]{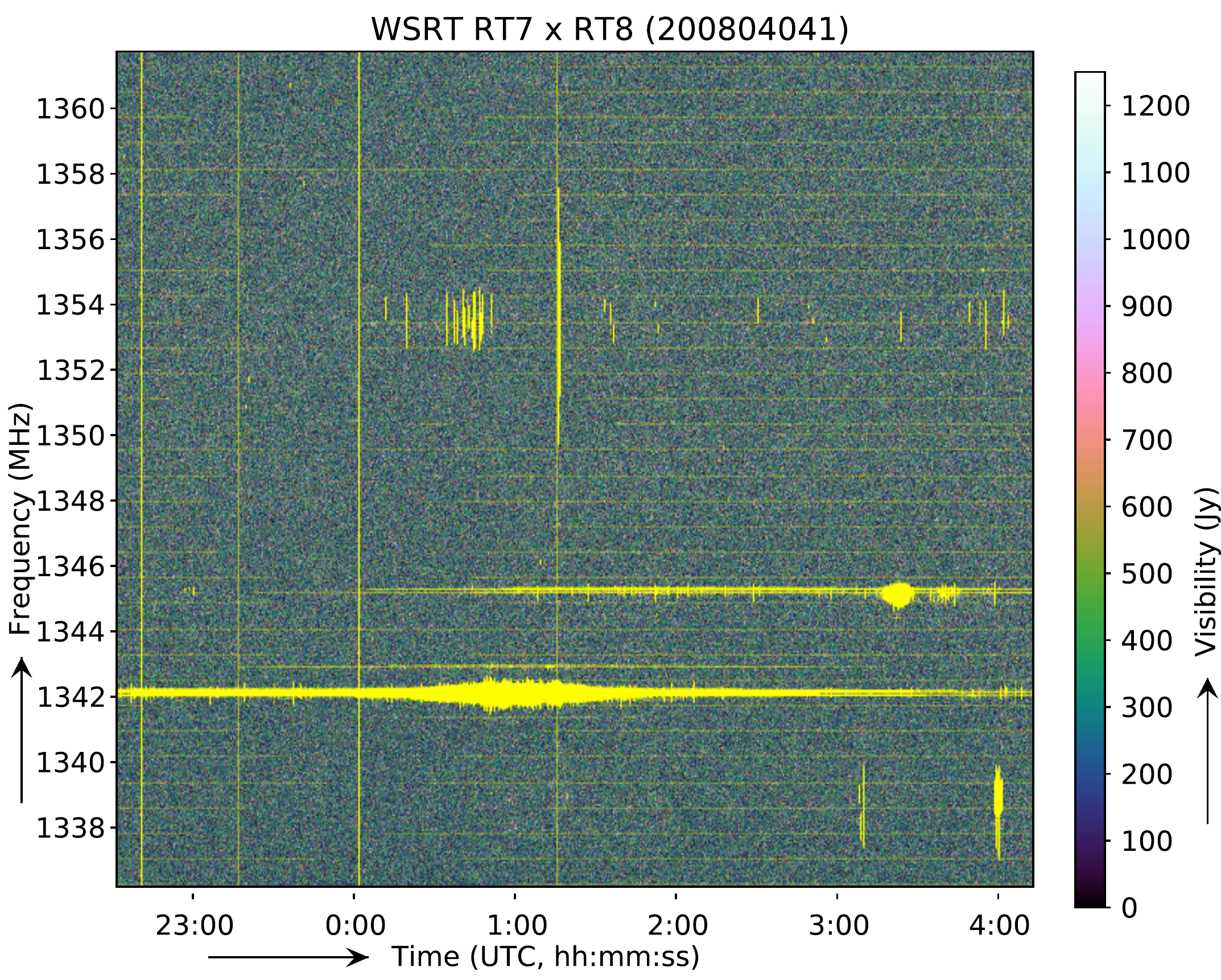}%
	\caption{Static sub-band band-pass correction before flagging with the Apertif flagging strategy. Top-left: input before correction; top-right: flagged without correction; bottom-left: input after correction; bottom-right: flagged with correction.}
	\label{fig:bandpass}
  \end{center}
\end{figure*}

\subsection{High-pass filtering} \label{sec:high-pass-filter}
The high-pass filter that is applied to remove astronomical source contribution before thresholding is, for computational reasons, implemented as a Gaussian low-pass filter followed by subtracting the difference between the input and the low-pass filtered result. The high frequency resolution of Apertif makes it necessary to use a large filtering kernel in the frequency direction. Effectively, a kernel with a Gaussian sigma of 875 channels and 2.5 timesteps is used. Before filtering, the data is averaged in the frequency direction by a factor of 175, and after low-pass filtering, the result is upscaled to the original resolution using nearest neighbour resampling. This allows a convolution with a much smaller kernel, improving the speed of this operation. The result is an approximate of a Gaussian high-pass filter, but for the purpose of removing the sky signal, this is sufficiently accurate.

\subsection{Bandpass correction} \label{sec:bandpass}

In the Apercal Apertif processing pipeline, the entire bandwidth of Apertif is used at once during RFI detection. This is different from the original LOFAR strategy, that flagged small (200~KHz) subbands independently. Using the entire bandwidth has the benefit that broadband RFI that covers several sub-bands can be detected. This is relevant for Apertif observations, which are affected by broadband transmitting satellites and radar.

Because the bandwidth of Apertif is subdivided into sub-bands using a poly-phase filter bank, the band shape of the poly-phase filter is imprinted on the data. An example of this is shown in the top-left panel of Fig.~\ref{fig:bandpass}. This is corrected for during calibration, but during flagging (which needs to be done before calibration) the shape is still present. 

Performing detection using the entire bandwidth but without correcting for the poly-phase filter bank causes sub-band edge channels to be flagged, because the edges cause sharp transitions that trigger the detector. Moreover, the deviations in the data caused by the band-edges decrease the sensitivity of the detection toward actual RFI. The top-right panel of Fig.~\ref{fig:bandpass} shows an example of flagging without bandpass correction.

To remedy this, we implement a sub-band band-pass correction step in the detector. This step corrects the poly-phase filter shape using a static, observation-independent correction. We determine the shape by performing gain-calibration on a clean region of the band, and average the solutions over the subbands. The bottom-left panel of Fig.~\ref{fig:bandpass} shows the resulting corrected data set, and the bottom-right panel of Fig.~\ref{fig:bandpass} shows the result of flagging the bandpass. As can be seen, the band-pass correction has decreased the number of false detections considerably. Some edge channels are still flagged, even after correction. This is caused by aliasing in the sub-band edge channels, which change the statistics of those edge channels slightly. This can lead to artefacts which are very similar to RFI, hence they are occasionally flagged. This flagging is normally of limited concern, because those sub-band edge channels that are flagged are of lower quality. Because of this, they are often discarded during imaging.

\begin{figure}
  \begin{center}
	\includegraphics[width=9cm]{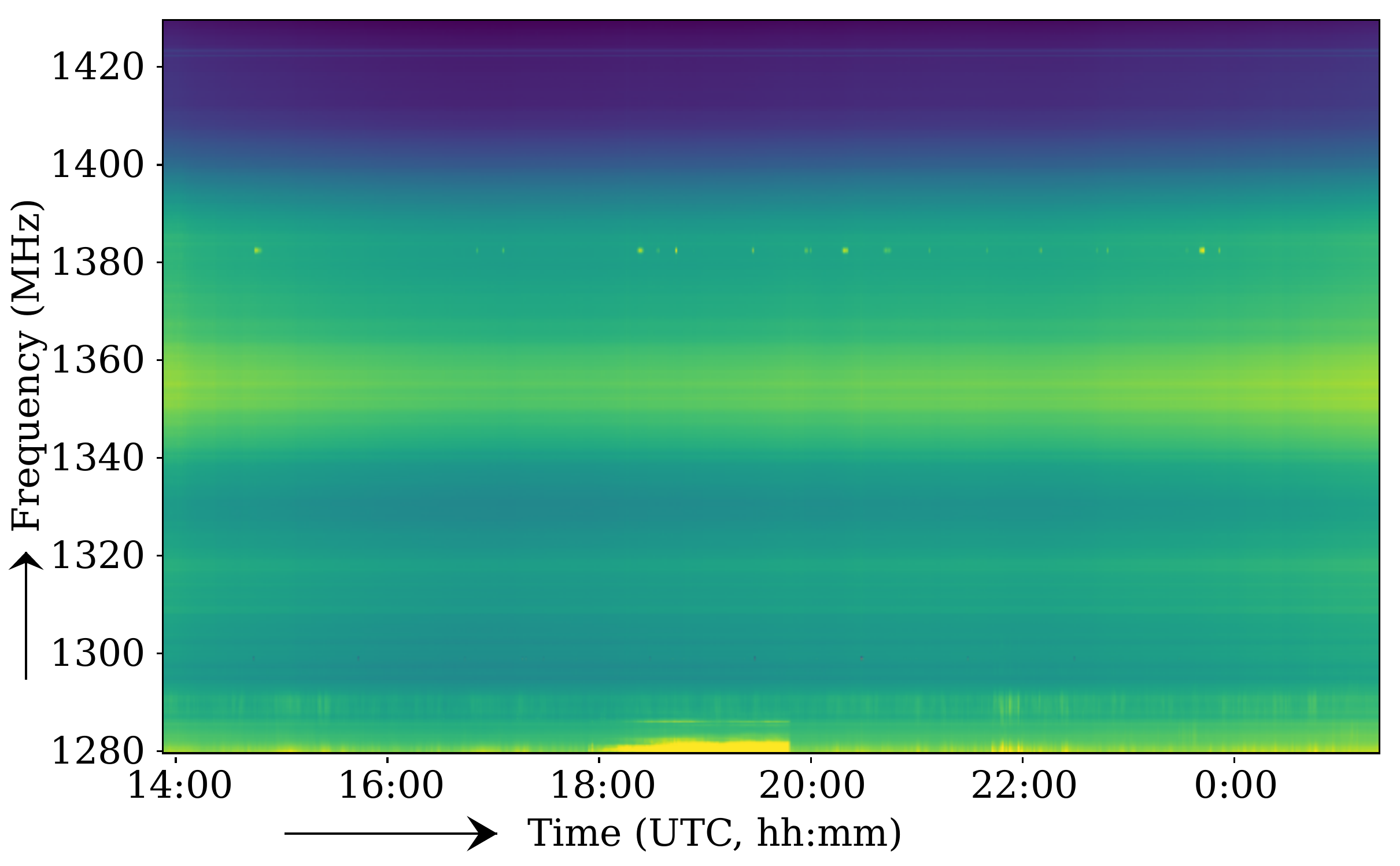}\\\includegraphics[width=9cm]{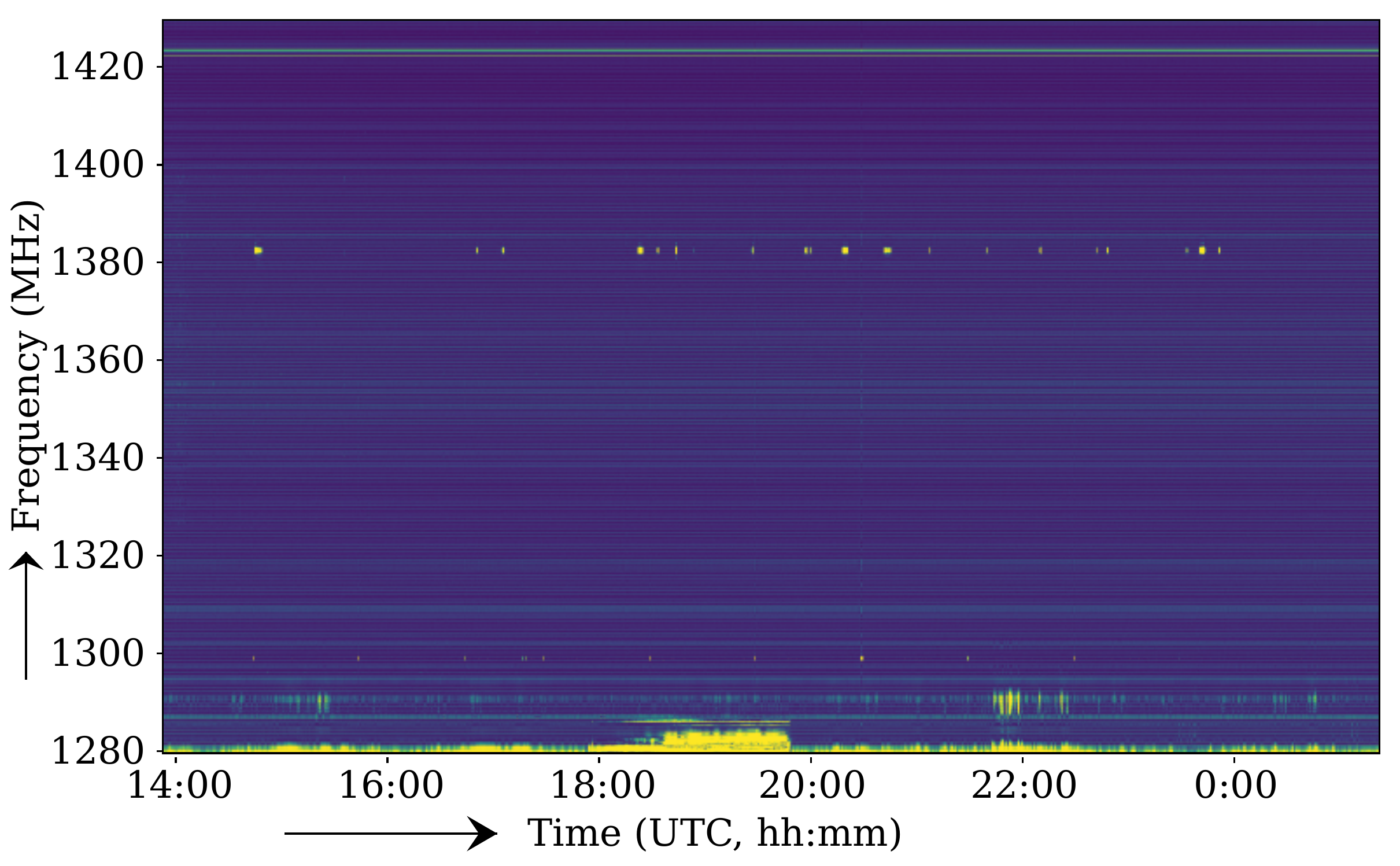}\\\includegraphics[width=9cm]{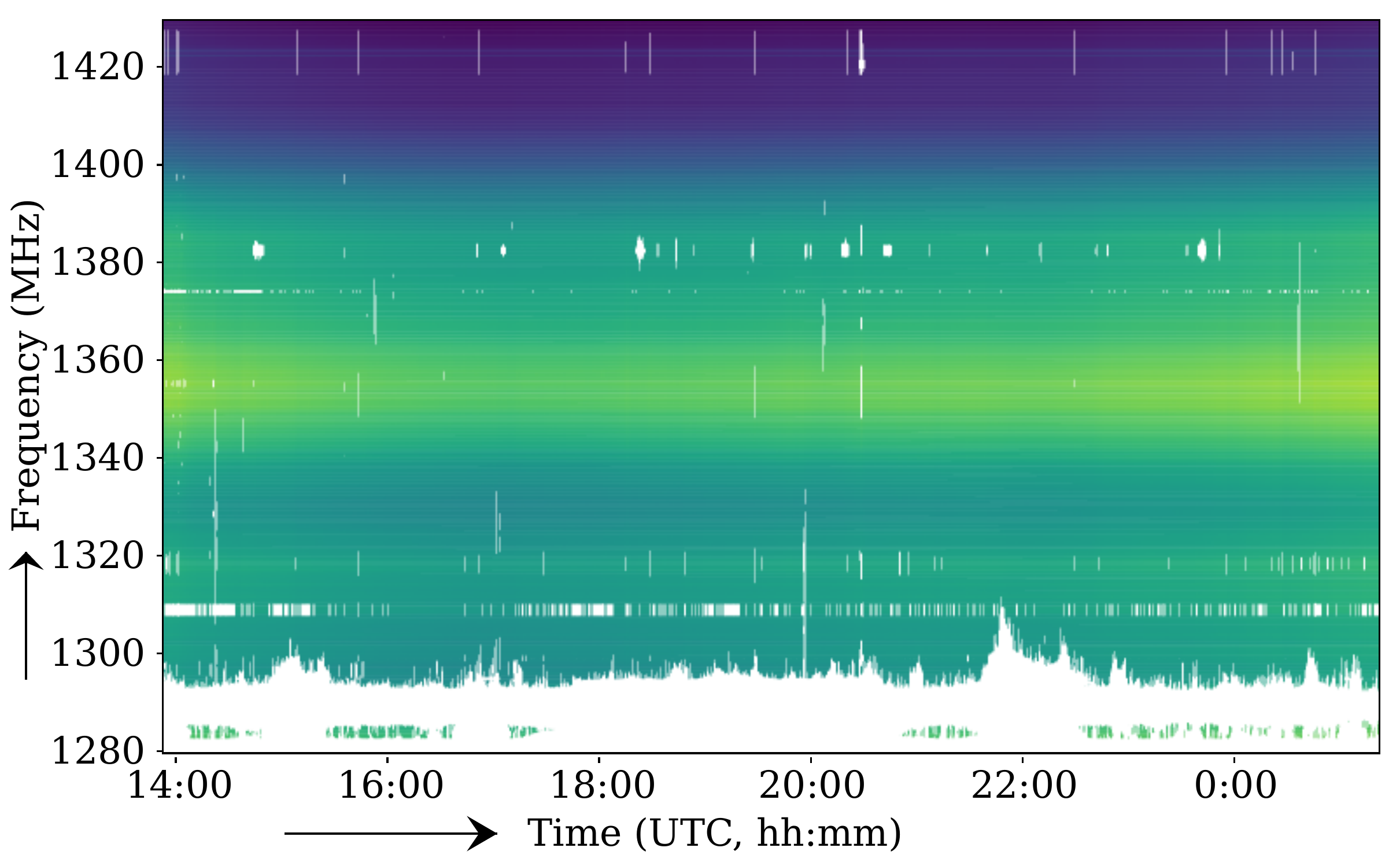}\\%
	\caption{Flagging of auto-correlations. Top image: input after sub-band band-pass correction; centre image: same after iterative high-pass filtering and with 10x more sensitive colour scale; bottom image: after flagging with the auto-correlation specific strategy. Because auto-correlations have different properties compared to cross-correlations, they require a specialized flagging strategy.}
	\label{fig:autocors}
  \end{center}
\end{figure}

\subsection{Flagging of auto-correlations}
Given the output voltage of the two feeds of the same antenna, $\mathbf{e} = \left( e_x, e_y \right)$, auto-correlated visibilities are formed by taking the product $\mathbf{e}^H\mathbf{e}$ (i.e., the outer product $\mathbf{e}\otimes\mathbf{e}$) and integrating, resulting in \texttt{XX}, \texttt{XY}, \texttt{YX} and \texttt{YY} visibilities. While auto-correlations are not often used for scientific data products, they are useful for system monitoring and quantifying the system noise. For such analyses, it is desirable to flag RFI.

Compared to cross-correlated visibilities, auto-correlated visibilities have different properties: in the \texttt{XX} and \texttt{YY} correlations, system noise and RFI will not decorrelate, and auto-correlated visibilities are sensitive to the global sky signal instead of fluctuations in the sky signal.

An example of auto-correlated dynamic spectrum from Apertif is shown in the top image of Fig.~\ref{fig:autocors} (after sub-band band-pass correction as described in \S\ref{sec:bandpass}). Compared to cross-correlations such as shown in Fig.~\ref{fig:bandpass}, the dynamic spectrum of auto-correlated visibilities appears much smoother, is systematically offset from zero and contains stronger structure in the frequency direction.

The flagging strategy that was optimized for the cross-correlations detects RFI by comparing high-passed filtered amplitudes of visibilities to the variance of these amplitudes. Because the amplitude variance is much lower compared to cross-correlations, this results in flagging auto-correlations with increased sensitivity. At the same time, the auto-correlations contain stronger instrumental frequency-structure. These two effects combined causes the cross-correlation flagging strategy to flag all of the visibilities of the auto-correlations of Fig.~\ref{fig:autocors}.

To solve this, we use a different flagging configuration for the auto-correlations. The difference with the cross-correlation strategy is as follows:
\begin{itemize}
 \item[-] The time-direction \textsc{sumthreshold} step (sensitive to consistently high values in the time direction, e.g. band-pass structure) is reduced in sensitivity by a factor of 6.
 \item[-] The frequency-direction \textsc{sumthreshold} step (sensitive to consistently high values in the frequency  direction, e.g. broad-band RFI) is reduced in sensitivity by a factor of 2.
 \item[-] The size of the high-pass filter kernel is reduced by 3.5 in the frequency direction, to filter out more of the spectral gain fluctuations of the instrument.
 \item[-] The number of iterations is increased from 3 to 5. This increases the required computations but improves robustness in the presence of a large dynamic range, as is the case for auto-correlations.
 \item[-] Only the \texttt{XX}, \texttt{XY} and \texttt{YY} correlations are used for detection, to reduce unnecessary computations. \texttt{YX} correlations are equal to the conjugated \texttt{XY} correlations, and using these for flagging does not provide additional information.
\end{itemize}

A result of this auto-correlations strategy is shown in the bottom image of Fig.~\ref{fig:autocors}. Visual inspection shows that all visible RFI is indeed detected, and the number of false detections appears low. Because we do not have a ground truth, we do not try to quantify these results. Similar to the cross-correlation strategy, the auto-correlation strategy flags parts of the sub-band edges. The centre image of Fig.~\ref{fig:autocors} shows the high-pass filtered data of the final iteration.

\begin{figure*}
  \begin{center}
	\includegraphics[width=9cm]{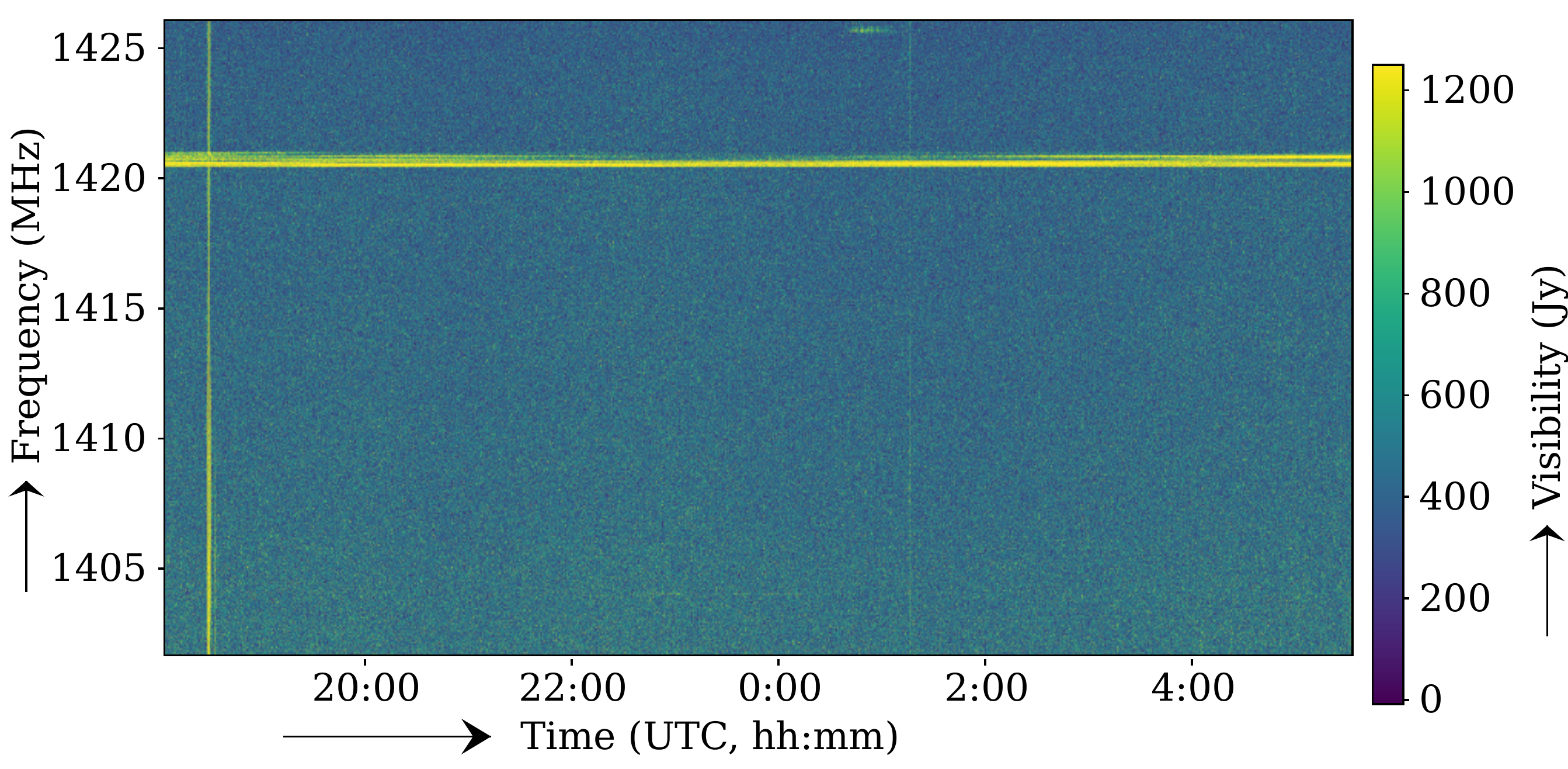}\hspace*{3mm}\includegraphics[width=9cm]{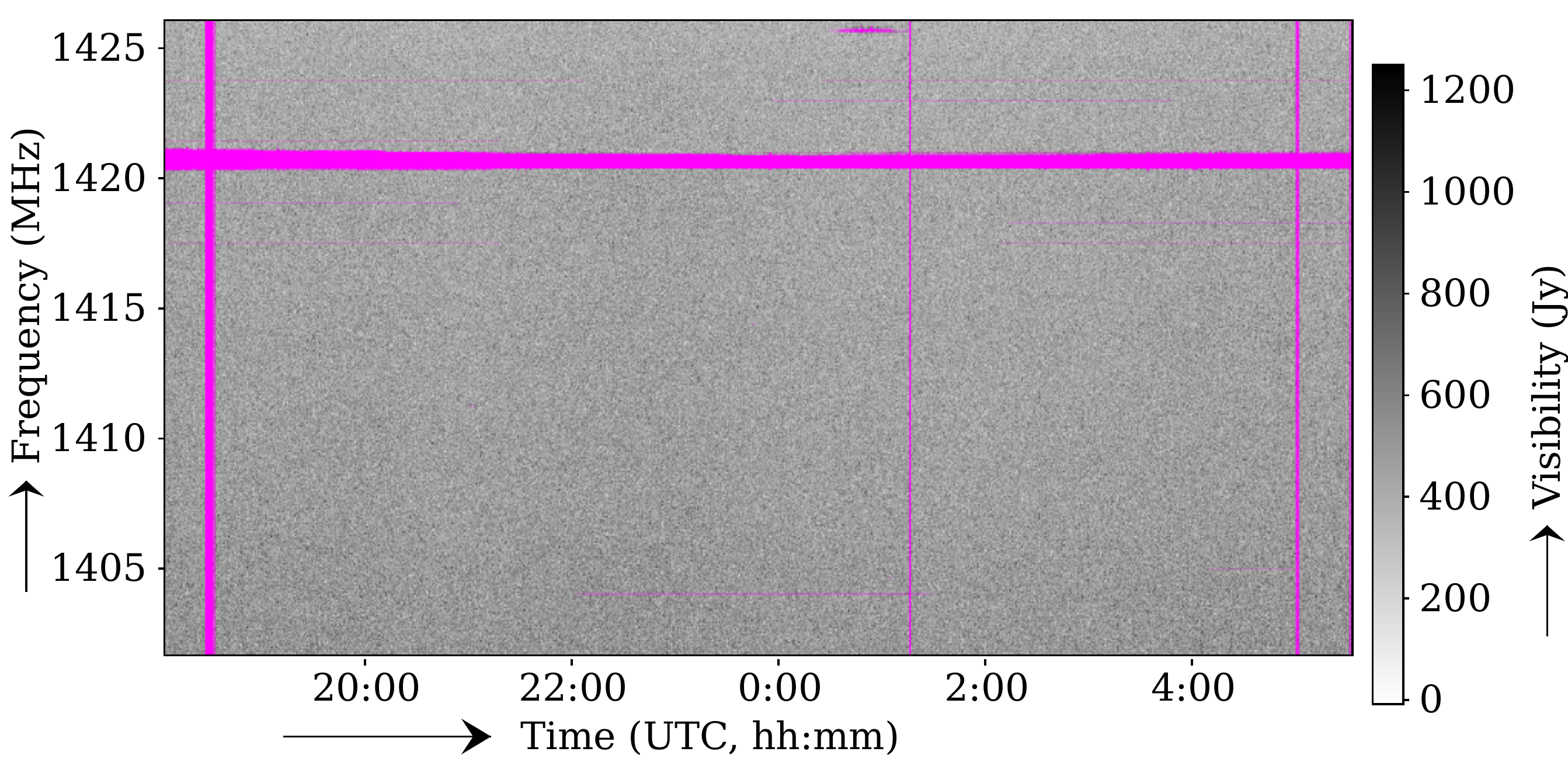}\\	\includegraphics[width=9cm]{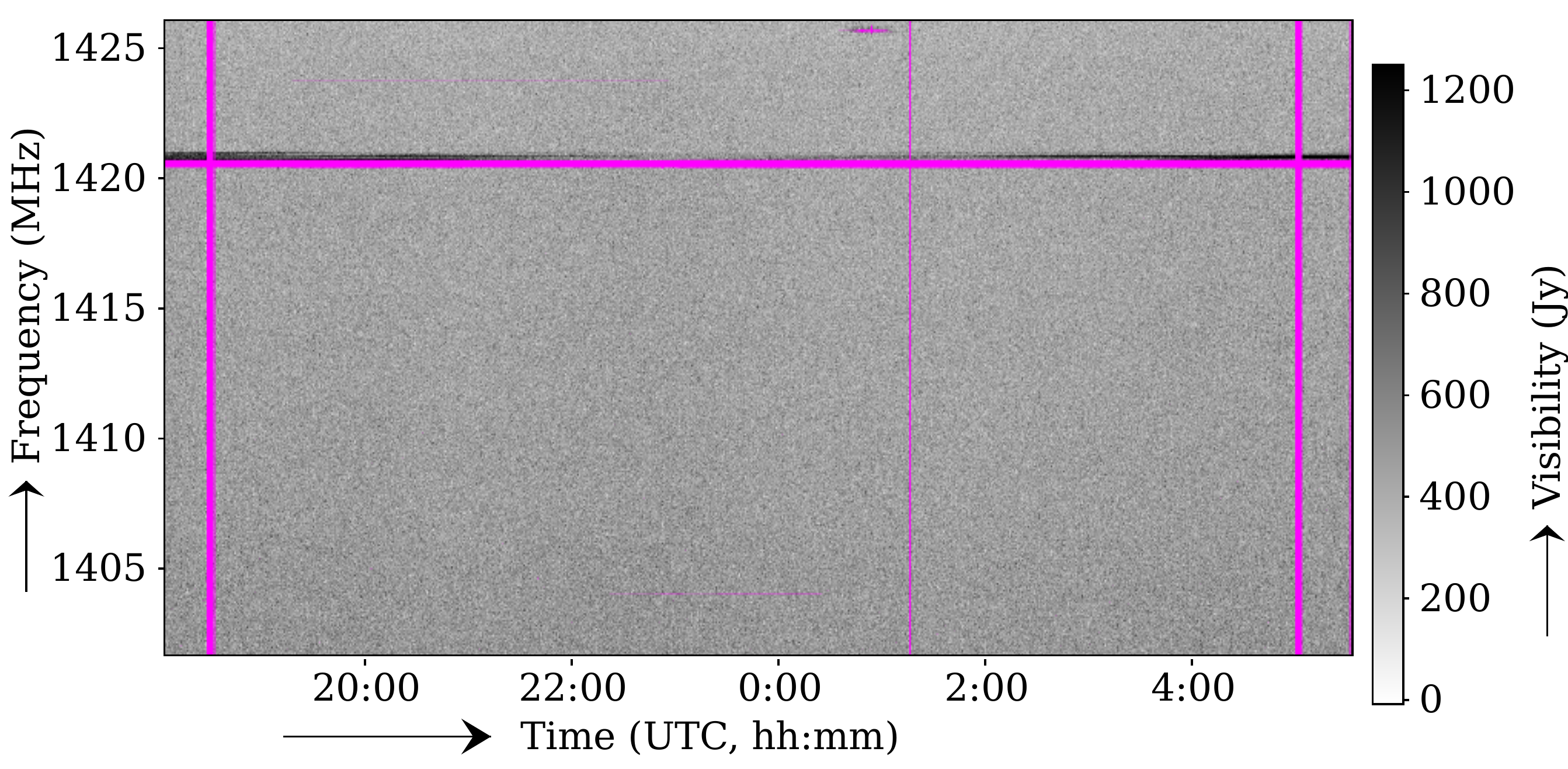}\hspace*{3mm}\includegraphics[width=9cm]{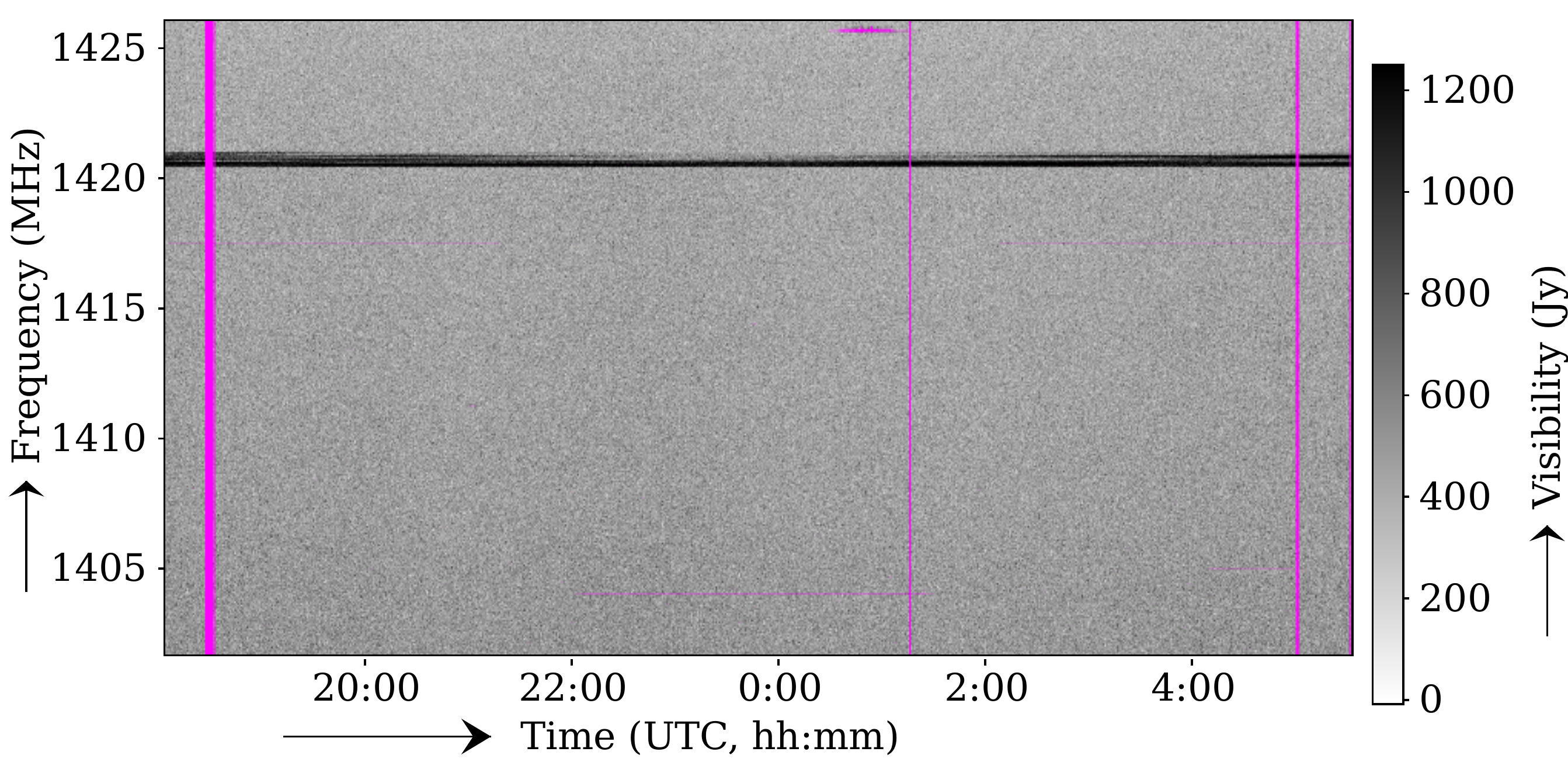}\\%
	\caption{Band-pass corrected M31 data from WSRT RT9 $\times$ RTA with a strong HI signal. Top-left image: input data. The bright emission around 1420~MHz is from HI and should not be flagged. The vertical lines are instrument or RFI artefacts that should be flagged. Top-right image: after RFI detection without HI modifications, showing in pink what is flagged. Bottom-left image: after RFI detection using Stokes Q, U and V. Bottom-right image: after RFI detection using a specialized strategy for 1418-1424 MHz.}
	\label{fig:hi-signal}
  \end{center}
\end{figure*}

\subsection{Avoiding HI removal} \label{sec:avoid-hi}
In observations that cover bright nearby galaxies or the Galactic plane, the 1420~MHz HI line may be detectable in the visibilities from a single cross-correlated baseline. For example, the top-left image of Fig.~\ref{fig:hi-signal} shows one baseline from a M31 observation, which clearly shows a contribution from HI-emission around 1420~MHz. This poses a challenge for RFI detection, because such a fine, spectrally-consistent signal is quite similar to RFI. As shown in the top-right image of Fig.~\ref{fig:hi-signal}, when standard flagging is performed on these data, the HI emission is detected as RFI.

We analyze different ways to mitigate this. In the Netherlands, frequencies between 1400-1427 MHz are reserved for radio astronomy and other forms of passive research\footnote{The Dutch spectrum allocations can be found at \url{https://www.agentschaptelecom.nl/}}, and transmitting inside this band is not allowed. As a result, these frequencies are almost free of man-made emission. A simple mitigation strategy is therefore to disable RFI detection inside this band. Unfortunately, the recorded visibilities do occasionally contain strong, non-astronomical values inside this band. The three vertical lines in the images of Fig.~\ref{fig:hi-signal} are an example of such an observation. Most frequently, these are caused by saturation of a receiver, causing a broadband-like signal in the recorded visibilities, although they might occasionally be caused by RFI emitted at these frequencies (e.g. from a sparking device or lightning). Leaving these broadband contaminants in the data causes degradation of the images. In particular, they cause visible stripes in continuum, full bandwidth images.

Another approach is to flag only based on Stokes Q, U and V. Man-made RFI is often polarized, whereas the sky emission in these polarizations is generally much fainter. The result of this approach is shown in the bottom-left image of Fig.~\ref{fig:hi-signal}. While a part of the HI emission has been left intact, it is still bright enough in these polarizations to get detected. This is even the case when flagging on only one of these polarizations: the HI emission is present in all of the polarizations. Moreover, we occasionally observe RFI that is only visible in Stokes I, and removing any of the polarizations decreases the effectiveness of RFI detection. In Fig.~\ref{fig:hi-signal}, the transmitter around 1425 MHz / 0:00 UTC is for example not as well detected in this approach compared to standard flagging.

Because none of these approaches give good results, we consider another approach, and run the flagger twice: in run A) we flag the data with the normal detection strategy, and in run B) we run the detection with a strategy that is insensitive to spectral lines. For frequencies outside the HI range we use the flags from run A), and inside the HI range (1418--1424 MHz) we use B). The result of this approach is shown in the bottom-right image of Fig.~\ref{fig:hi-signal}. With this approach, broadband structures have been detected as RFI and HI emission is left in the data.

To avoid flagging spectral lines in run B), we adjust the following flagging settings during this run):
\begin{itemize}
 \item[-] The high-pass filter in frequency direction is set to have a kernel size of one channel, to filter out fluctuations in frequency.
 \item[-] The sensitivity of the time-direction sumthreshold step is decreased by a factor of 4, to reduce flagging of line-like structures.
 \item[-] The sensitivity of the frequency-direction sumthreshold step is decreased by a factor of 2. This reduces flagging of temporal fringes in HI emission.
 \item[-] The number of iterations is increased to remain robust in the presence of strong HI emission.
\end{itemize}
On overall, the resulting strategy is almost entirely insensitive to spectral-line-like structures. The sensitivity to broadband structures will also be reduced because of these changes, but given that this strategy remains sensitive to faint broadband structures such as shown in Fig.~\ref{fig:hi-signal}, we consider this tolerable.

Because run B) requires only a small part of the full bandwidth, the second flagging run is relatively fast, hence the increase in computations caused by this is modest (about 20\%).

\subsection{Reading overhead and memory considerations}
During the AOFlagger stage of the \textsc{apercal} pipeline, observations are stored in the Casacore Measurement Set format. In this format, the data of an observation is lexicographically sorted in time, and then in baseline and frequency. While this ordering is suitable for calibration, flagging requires the data baseline by baseline. Unfortunately, the data for a single baseline is spread throughout the file. Therefore, reading a baseline requires reading the file from beginning to end. Because of the block size and caching of storage media, it is inefficient to read the baselines one by one with this approach.

AOFlagger supports three methods for accessing the data:
\begin{itemize}
 \item [-] Direct reading. In this mode, the data is directly read from the measurement set just before they are needed. Because multiple baselines are processed in parallel using multi-threading, a few baselines are read from the measurement set at once. This mode results in scanning through the input data multiple times, which is computationally costly.
 \item [-] Reorder before processing. In this mode, the whole measurement set is reordered by baseline, frequency and then time and rewritten to disk in a binary, internal format before processing is started. This results in reading the data only twice and is generally faster than the direct reading mode, but requires disk space to store the copy of the data.
 \item [-] In-memory data. In this mode, the whole measurement set is read into memory before starting processing. This results in reading the data only once and is generally the fastest mode, but requires a considerable amount of memory.
\end{itemize}
Apertif data sets are large and expensive to read: reading the data more than once is undesirable. As a result, the only acceptable reading mode is the in-memory mode. In the particular computing mode where Apercal runs, the amount of memory required by this mode is a considerable constraint, and requires a dedicated node for each flagging operation performed. 

Other observatories have solved this issue by integrating \textsc{aoflagger} into a multi-step preprocessing pipeline that stream through the data, split the data in time for flagging and hand these data over part by part to AOFlagger via its application programming interface. Examples of such pipelines are \textsc{cotter} \citep{offringa-2015-mwa-rfi} and \textsc{DP3} \citep{vandiepen-dp3}, which are preprocessing pipelines for the Murchison Widefield Array and the Low-Frequency Array, respectively. In this approach, several tasks (e.g. conversion, phase rotation, flagging, averaging, compression) can be applied with a single read through the data, thereby reducing the read overhead. In the case of Apertif, such a streaming pipeline does not exist. Instead, aoflagger runs as a stand-alone tool inside Apercal.

To solve the memory and reading issue for Apertif, we implemented a time-chunking approach into aoflagger. In this mode, aoflagger reads small chunks in time and flags these independently. This makes it possible to use the memory reading mode, because the data for individual chunks is small enough to fit in memory. It does imply that the algorithm has less information available to do its RFI detection. Therefore, it is important to let time chunks still have a significant size, because AOFlagger would otherwise not be able to find faint RFI, that is persistent in time, but not detectable in a small chunk. For Apertif, we use a chunk size corresponding to about half an hour of data.

\subsection{Use of Lua}
Before AOFlagger version 3, AOFlagger strategies were written in the 
extensible markup language (XML). An XML file specifies a sequence of steps and is interpreted by AOFlagger, and this sequence is executed separately for the data from every baseline. The sequences run multi-threaded, and reading and writing of data is done outside of the strategy. Examples of XML steps are to calculate visibility amplitudes; running \textsc{sumthreshold} or \textsc{sir} operations on the data; or to combine the flags of all polarizations.

Over the years, the use of AOFlagger extended to more and more use-cases: different telescopes, flagging after calibration, high-resolution flagging, etc. It became desirable to make the strategies more flexible. In particularly, it became desirable to support standard scripting structures such as loops, conditionals, variables and to provide standardized documentation of the steps. The idea was therefore formed to embed a standard interpreter into AOFlagger and provide a function interface for each step. The data-intensive computations are still performed by high-performance precompiled C++ code, while these are glued together using an interpreted script, thereby combining flexibility with high performance.

Our first approach was to embed it into Python, because of its popularity in astronomical data science. After having implemented a prototype that embeds the Python interpreter into AOFlagger, it turns out some of the features of the Python interpreter conflict with how AOFlagger runs these scripts. Particular challenges were to deal with the global interpret lock; memory management; and fast restarts of the interpreter. While there are various ways to work around these issues, the design goals of the Python language and interpreters do not focus specifically to make the language embeddable.

Lua\footnote{https://www.lua.org/} is a scripting language that is widely used for embedding scripts in applications, notably in computer games to implement scripted game sequences. This scenario is close to the AOFlagger use-case: the interpreter is integrated into such games, called many times and supports multi-threaded script execution. Algorithmic code that requires high performance can be implemented in compiled languages (C++ in the AOFlagger case). With this idea in mind, we decided to integrate the Lua interpreter into AOFlagger and implement all steps as Lua functions.

The use of a full scripting language has increased the possibilities inside the flagging strategies considerably. For example, it is now possible to adapt the strategy based on properties such as the baseline length, frequency, auto- or cross-correlation, etc. A consequence of the new interface is that existing strategies need to be rewritten, which can not be done automatically. All default strategies have been rewritten to use Lua, which currently includes specialized scripts for 11 observatories (Aartfaac, APERTIF, Arecibo, ATCA, Bighorns, JVLA, MWA, WSRT, LOFAR, NenuFAR). These have all been verified to produce the same result as the old XML-based strategies. Because the new function interface gives better control over what steps need to be run, the speed of the new strategies is slightly higher (several percent). We do not notice any significant overhead from using Lua: the computational time is dominated by the computations inside the function calls.

\section{Results} \label{sec:results}
Apertif observations are processed by the automated Apercal pipeline. This pipeline includes the flagging strategy as described in Sec.~\ref{sec:method}. In this section, we present results of the full flagging step on Apertif observations. The data that we look at has been recorded between 2019 and 2022. Science products from the first year of observing have been described in the first Apertif data release \citep{adams-2022, kutkin-2022}.

\subsection{RFI detection examples}

\begin{figure*}[hbtp]
  \begin{center}
	\includegraphics[width=18cm]{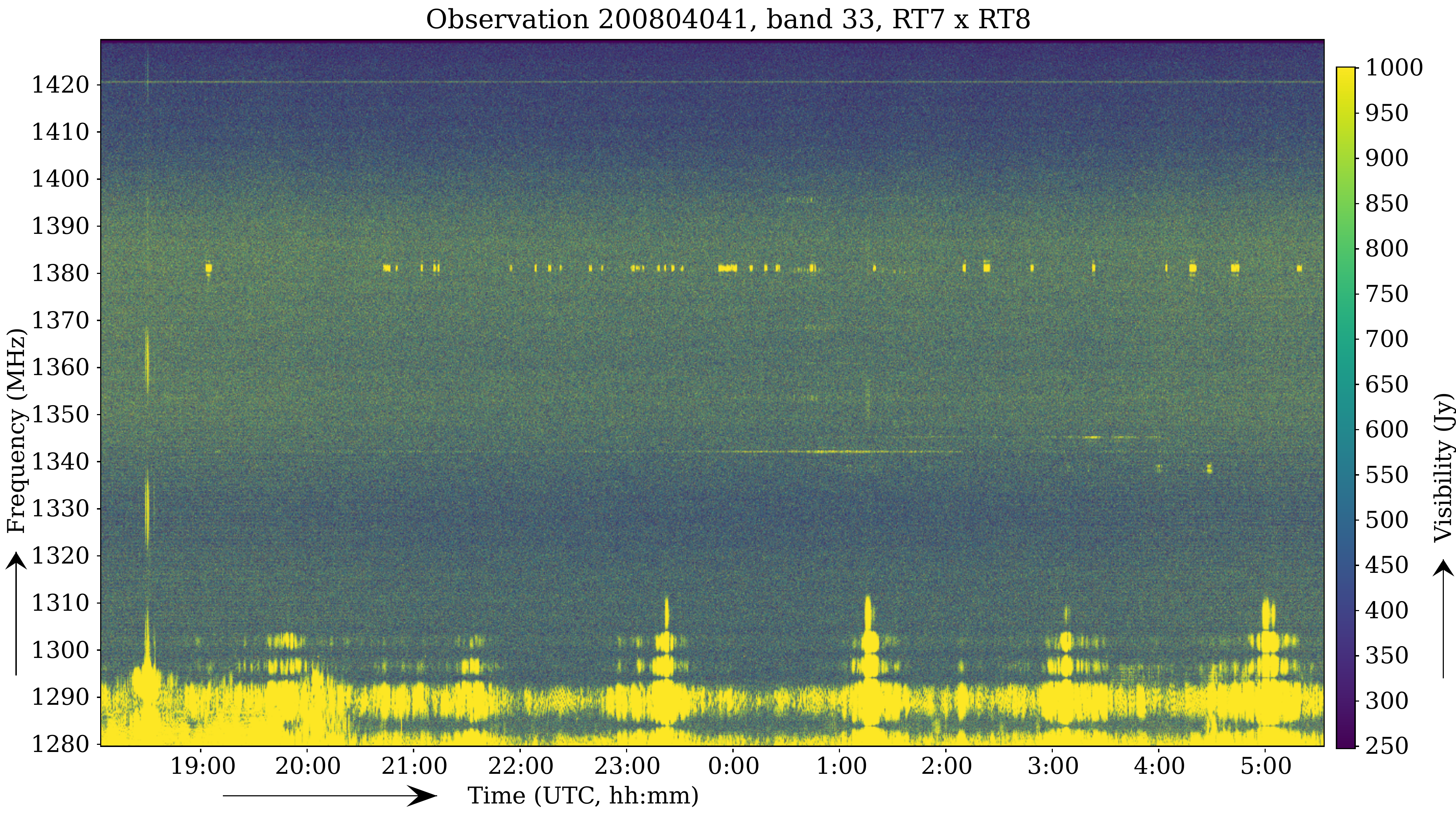}\\%
	\includegraphics[width=18cm]{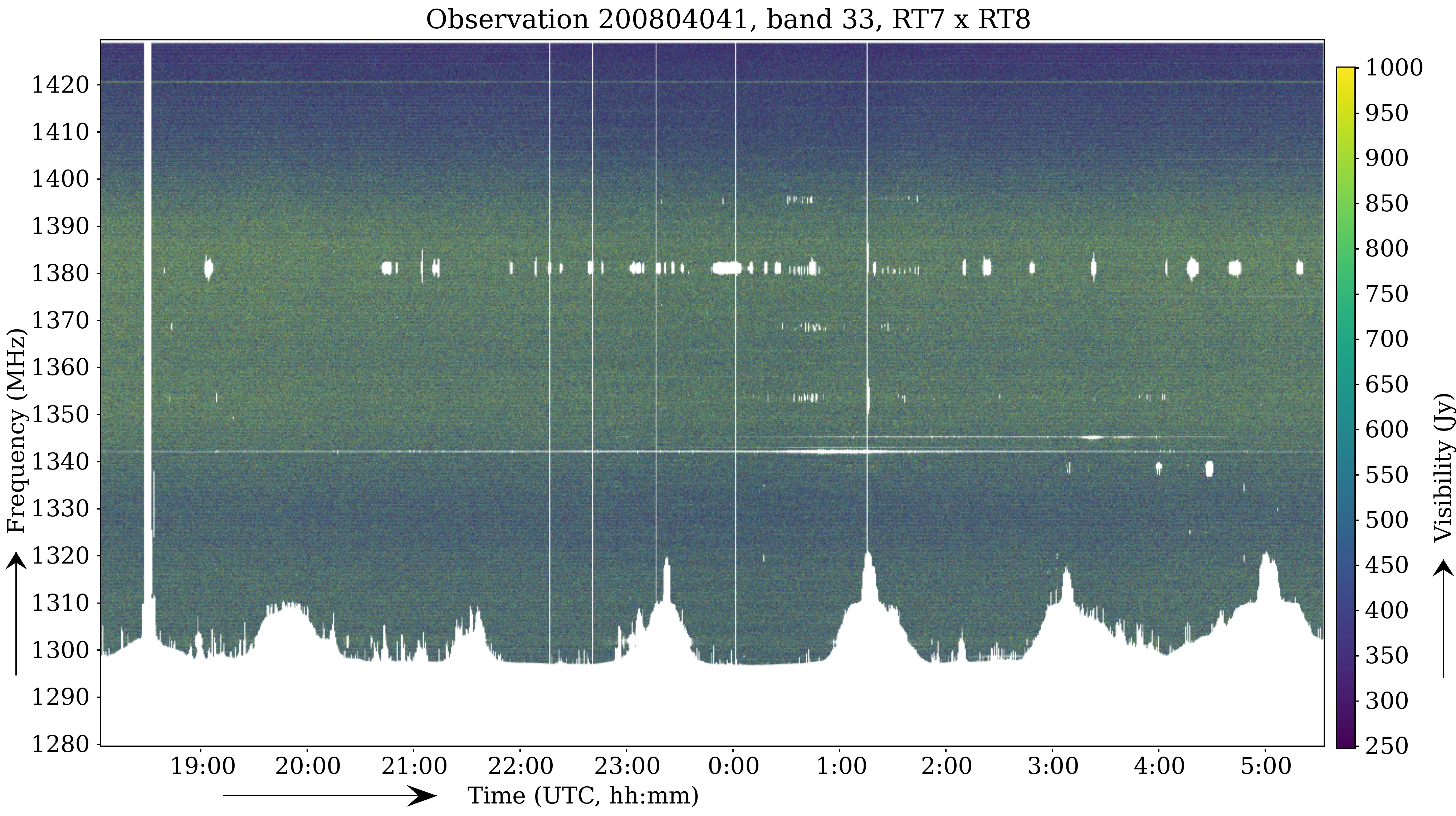}%
	\caption{Typical flagging result for a single baseline in a wideband observation. The top panel shows the input visibilities, and the bottom panel shows the visibilities overlaid with the detection result in white. These plots show the Stokes I visibilities. Some interference features are only visible in Stokes Q, U or V, such as the vertical features around midnight. All interference features have successfully been detected, and no obvious undesirable detections are visible, with the exception of horizontal flagged features every 200~kHz, caused by the sub-band bandpass (see Fig.~\ref{fig:bandpass}). 18\% of the data gets flagged for the baseline in this observation.}
	\label{fig:full-example-quiet}
  \end{center}
\end{figure*}

The detection strategy described in Sec.~\ref{sec:method} runs fully automated, and does not require further flagging before calibration and continuum imaging. In general, manual inspection of data after RFI detection shows no residual RFI and few false positives. Fig.~\ref{fig:full-example-quiet} shows the $1280$--$1430$~MHz range of a typical observation. The top plot shows the data before RFI detection, and the bottom plot shows in white what has been detected as RFI. Fig.~\ref{fig:full-example-loud} shows a challenging case with wider bandwidth, with a moderate amount of RFI, missing data (1200--1220~MHz) and strong fringes. Top and bottom plots show again before and after detection. This also demonstrates the challenging situation for radio astronomical science between 1150 and 1300~MHz. 

For continuum imaging, it is often useful (or at least pragmatic) to take out any visibility that appears to have a contribution from RFI. For spectral imaging, a flagging result such as shown in Fig.~\ref{fig:full-example-loud} is problematic, because many channels are fully removed. In those cases, it is possible to reduce the sensitivity of the RFI detection. The sensitivity is specified as a variable in the script. For the detection result shown in Fig.~\ref{fig:full-example-less-sensitive}, the sensitivity was decreased by a factor of 3. Compared with the result in Fig.~\ref{fig:full-example-loud}, this reduced the flagging from 49\% to 33\%. This takes out the strongest RFI, but leaves weak (but visible) RFI in the data. Decreasing the sensitivity further continues to trade the availability of visibilities with a lower quality of those visibility.

\subsection{RFI characteristics and long-term statistics}
\begin{figure*}[h]
  \begin{center}
	\includegraphics[width=14cm]{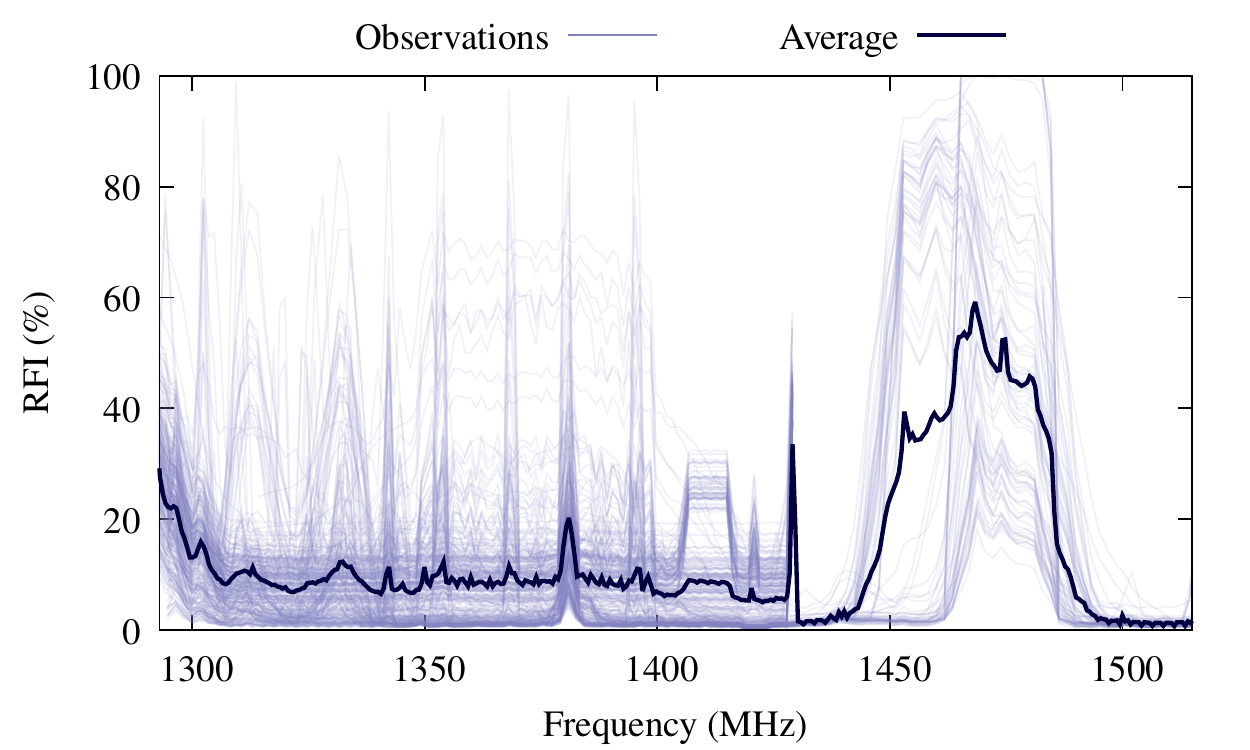}%
	\caption{Percentage of RFI over frequency detected in 304 Apertif observations, excluding auto-correlations.}
	\label{fig:rfi-percentage-spectra}
  \end{center}
\end{figure*}

During the flagging step, statistics are collected that summarize the (detected) RFI occupancy and data quality. We have collected these statistics for 304 of the currently processed observations. Averaged over all these observations and the full bandwidth, the total detected RFI occupancy is 11.1\% in the cross-correlated baselines and 14.6\% in auto-correlated baselines. Fig.~\ref{fig:rfi-percentage-spectra} shows the detected spectral RFI occupancy for each observation, as well as the occupancy averaged over all observations. Only cross-correlated data is included. At most frequencies, the average loss of data due to RFI is about 10\%, but with a spread of approximately 0-15\% between observations, and a few larger outliers. 

Frequencies between 1400 and 1427 MHz are reserved for radio astronomy. At these frequencies, the average RFI occupancy is slightly lower (approximately 8\%), but is evidently still affected by instrumental effects (such as receiver saturation) or natural and unintended RFI (such as lightning). Fig.~\ref{fig:hi-signal} shows data that is affected by such broadband artefacts. It is likely that the $\sim$10\% base-level of occupancy is caused by such artefacts.

Some observations show a small excess RFI occupancy at 1420~MHz. This is caused by HI that is detected as RFI. The methods to avoid flagging HI that are described in \S\ref{sec:avoid-hi} were implemented only halfway 2021. Some of the observations that are flagged before that still show false-positive detections at HI frequencies, but all observations after avoiding HI was implemented show indeed no HI flagging.

The same base level of 10\% is not visible at frequencies above 1430~MHz. The reason for this difference is that only a relative small number of observations cover frequencies above 1430~MHz. Frequencies between 1427 and 1492~MHz are allocated to various services, including mobile communication and fixed transmissions\footnote{See \url{https://www.agentschaptelecom.nl/}}. Some of these are satellite based. In 2020, the 1452---1492~MHz band was auctioned in the Netherlands and thereafter allocated for the use of 5G mobile phone downlink. As shown in Fig.~\ref{fig:rfi-percentage-spectra}, the use of data above 1430 MHz is limited.

Some channels between 1300--1400~MHz contain a few outlier RFI occupancies. These are caused by a nearby radar station that is occasionally turned on. Frequencies between 1130 and 1300\,MHz are predominantly affected by RFI from
Global Navigation Satellite Systems (GNSS), such as the US GPS,
Russian GLONASS, Chinese BeiDou, and European Galileo satellite
constellations. All these constellations use satellites in orbits at
$\sim$$2000$\,km and with high orbital inclinations ($i=54$--$65\degr$)
to provide global coverage. Frequencies for wide band transmissions
are assigned to, and shared between, these systems at 1176.45,
1191.795, 1207.14, 1227.6, 1278.75\,MHz (for GPS, BeiDou, Galileo) and
1202.025 and 1242.9375--1251.6875\,MHz (for GLONASS).

\begin{figure*}[hbtp]
  \begin{center}
	\includegraphics[width=18cm]{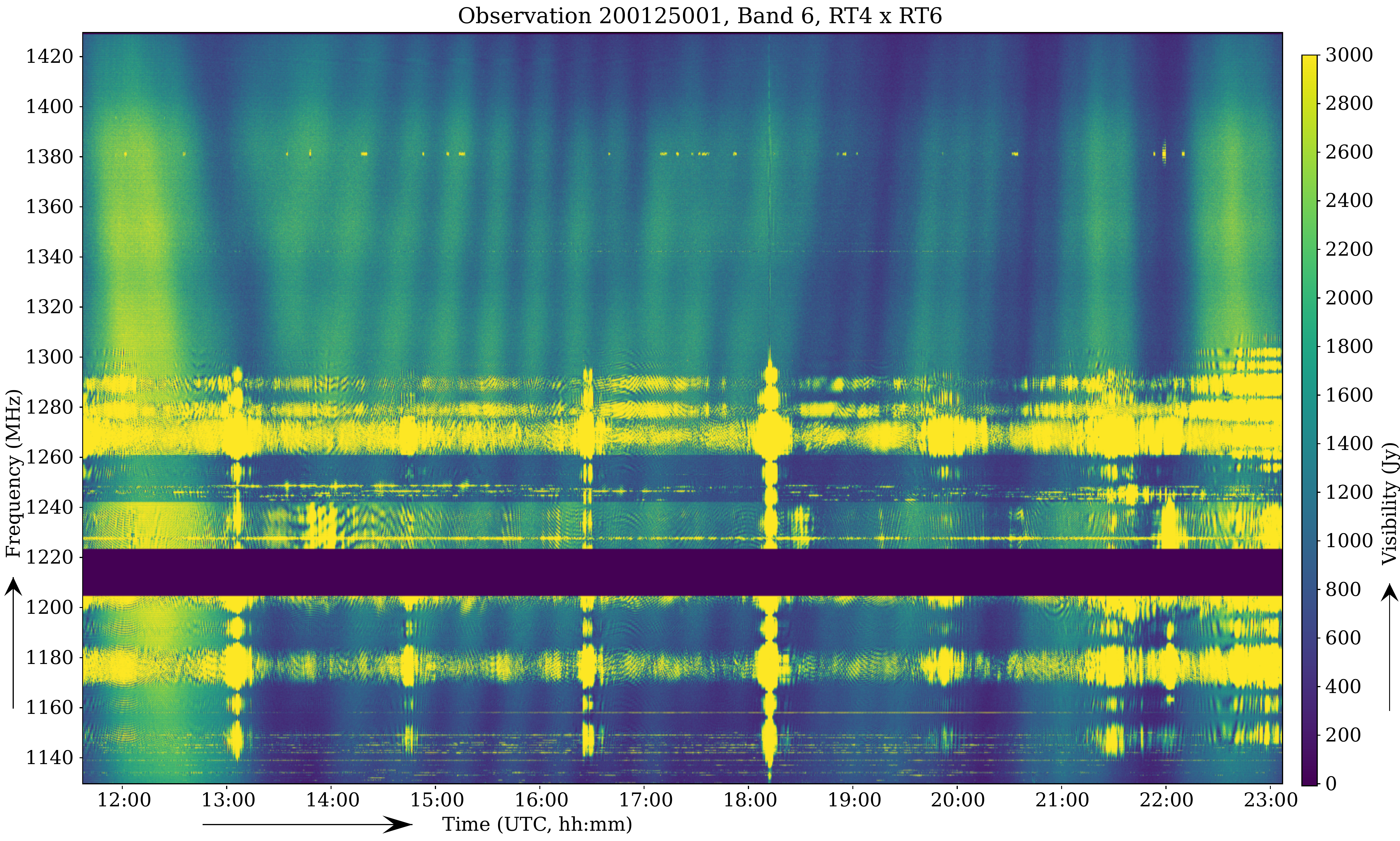}\\%
	\includegraphics[width=18cm]{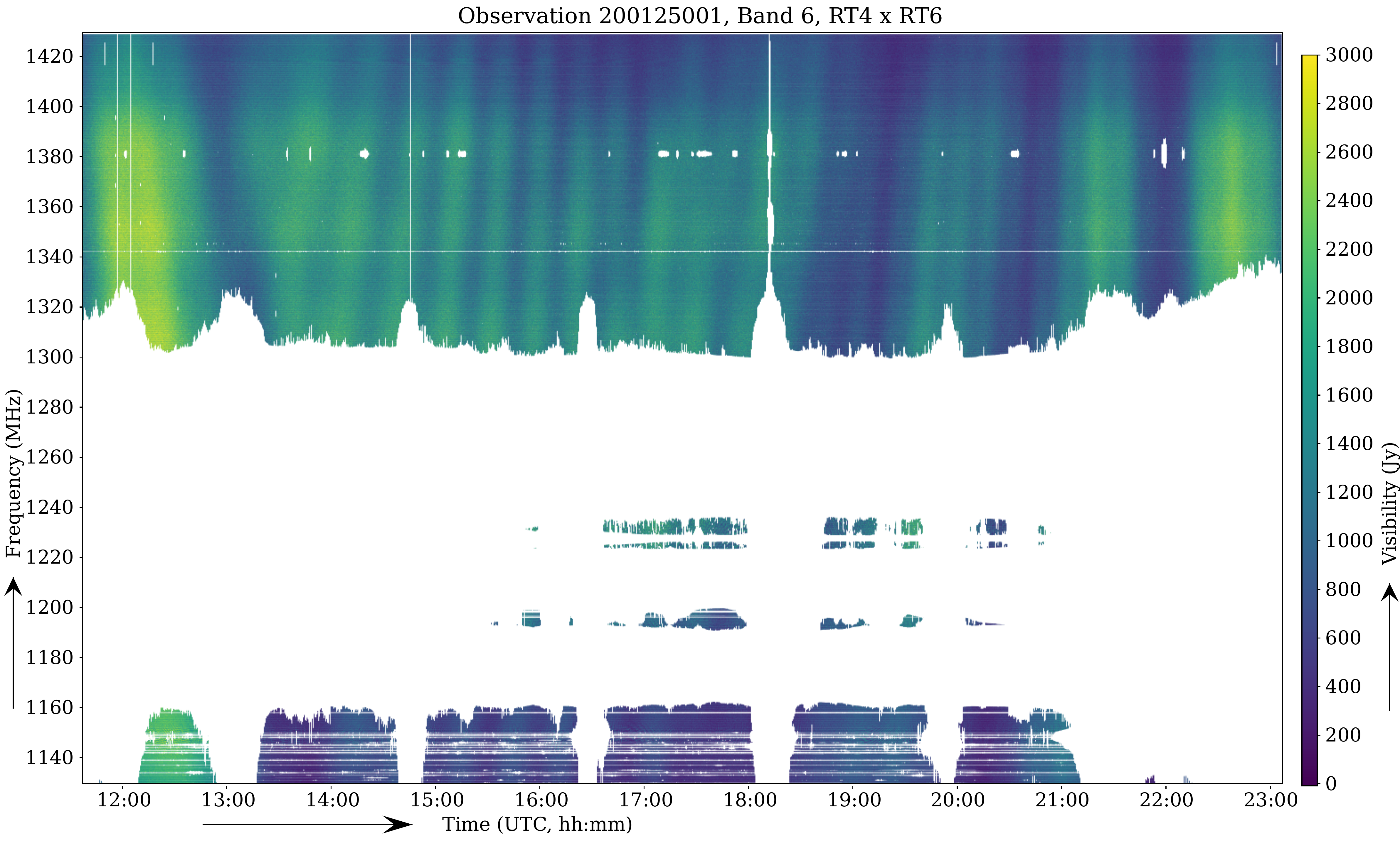}%
	\caption{Detection result for a full 300-MHz bandwidth observation.}
	\label{fig:full-example-loud}
  \end{center}
\end{figure*}

\begin{figure*}[hbtp]
  \begin{center}
	\includegraphics[width=18cm]{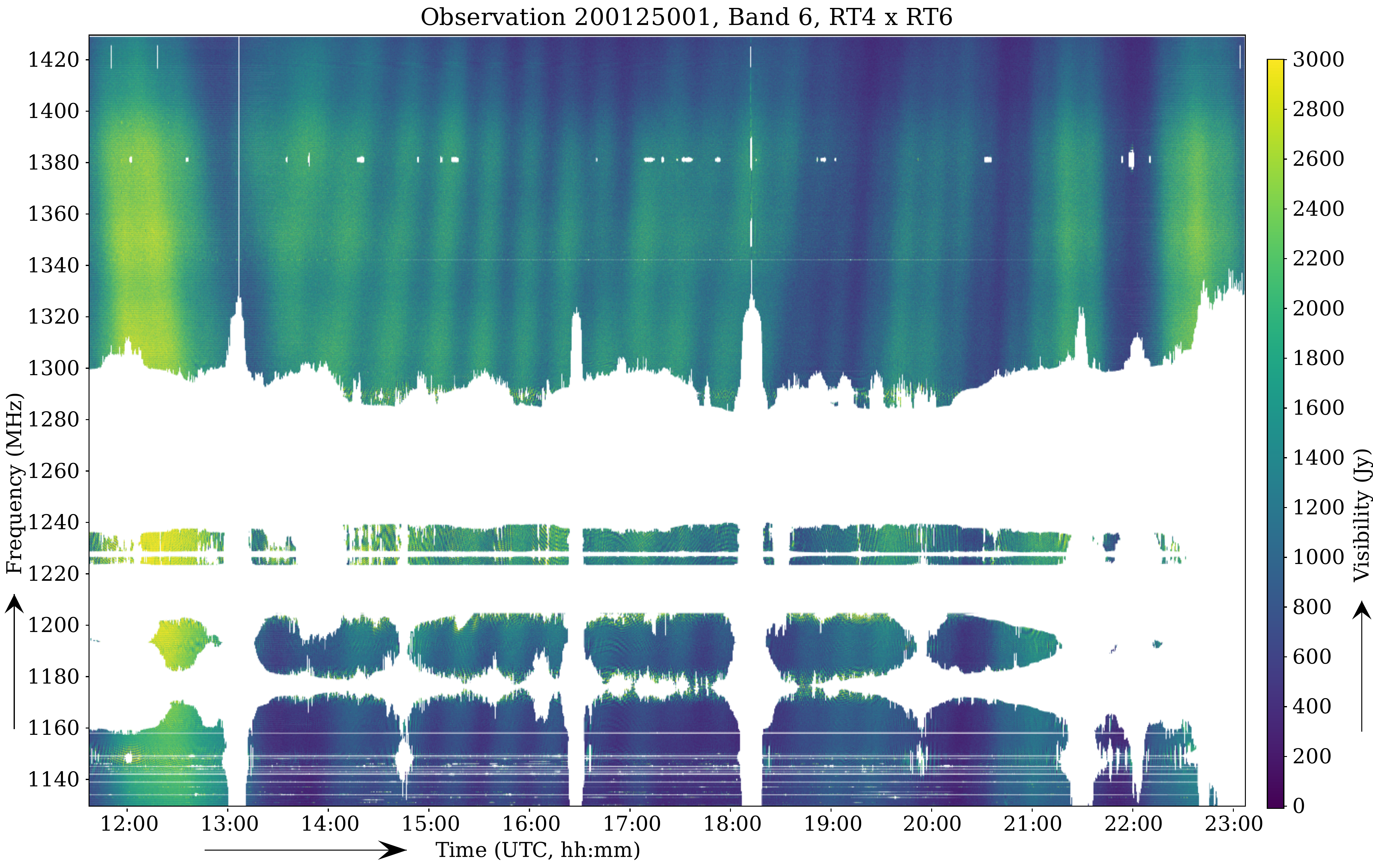}\\%
	\caption{Same as Fig.~\ref{fig:full-example-loud}, but flagged with 3$\times$ lower sensitivity.}
	\label{fig:full-example-less-sensitive}
  \end{center}
\end{figure*}

Wide band signals are detected at these frequencies throughout the
entire observation of Fig.~\ref{fig:full-example-loud} covering the band down to 1130\,MHz.
Using orbital ephemerides of these satellite constellations, we find
that the strong temporal RFI observed in Fig.~\ref{fig:full-example-loud} at 13:06, 14:46,
16:29, 18:13 and 19:54UTC is caused by BeiDou satellites passing
within $5\degr$ from the pointing of the APERTIF compound beam. The
pass of 18:13UTC had a minimum separation of $0\fdg31$ and led to
saturation of the receiver, affecting the entire observing band. Two
GPS satellites passed at $1\fdg47$ and $2\fdg30$ separation from the
beam pointing at 22:02 and 23:02UTC, and one Galileo satellite at
$3\fdg72$ at 22:59UTC, and coincident increases of the RFI levels are
observed, but not as strong as with the passes of BeiDou
satellites. The GNSS signals observed away from these passes near the
primary APERTIF beam are likely due to far sidelobes or multi-path
reflections of GNSS signals from the WSRT focus structure or other
nearby structures directly into the receiver.

\subsection{Computational requirements} \label{sec:speed}

In this section we summarize the computational requirements of the Apertif RFI detection strategy, with the aim of making it possible to approximate the computational requirements for other telescopes when a similar flagging strategy is used. Since the total throughput is depending on many complex factors of the computing platform (e.g. clock speed, cores, memory bandwidth, instruction set, vectorization), we aim at giving a first-order estimate only.

We measure the performance of flagging a set with visibilities from a single observation. We use an Apertif observation with 1346 timesteps, 24572 channels and 4 polarizations, for a total of 132M visibilities. This makes the visibility data, which consists of 4-byte single-precision real and imaginary values, 1.1~GB in size.

We perform our test on a desktop machine with an AMD Ryzen 7 2700X 8-Core processor and 64 GB of memory. This processor can perform hyper-threading, and thus we run 16 detections in parallel. We load the data in memory before detection and do not store the results, to avoid any disk access. Averaged over 10 runs, it takes 46 seconds to run 16 detections, which amounts to a throughput of 370 MB/s (or 46M visibilities/second). At the time of writing, a typical fast spinning disk achieves a sustained reading throughput of a few hundred MB/s. Hence, disk access can be a significant cost of a stand-alone RFI detection step. This can be problematic for supercomputers, because they have high computing power, but not a high I/O throughput.

\subsection{Comparison against a machine learning approach}
Some studies have found that machine learning can improve the accuracy of RFI detection. In \citet{nn-rfi-detection-2020}, the authors test their own \textsc{sumthreshold} implementation against a machine learning approach, using a ground truth flag mask that is manually determined by an engineer. Such a ground truth mask is difficult to make in general, including for Apertif data, where broadband RFI tapers off and it is unclear from which points samples are truly unaffected by RFI. We can however conclude that, after our pipeline, all visibly affected samples have been identified. Moreover, imaging results have achieved the thermal noise of the instrument, thereby indicating that the accuracy of interference detection is not a limitation.

This conflicts somewhat with the conclusions made by \citet{nn-rfi-detection-2020}. The \textsc{sumthreshold} implementation that is used there to compare their results with, does not achieve the published accuracy of \textsc{aoflagger}, because residual interference is visually present. Potential explanations for these differences could be i) that \citeauthor{nn-rfi-detection-2020} train their network for a specific scenario but did not optimize their \textsc{sumthreshold} approach; or ii) that they do not use a full (i.e. \textsc{aoflagger}-like) \textsc{sumthreshold}-based pipeline that includes the \textsc{sir} operation and that is similarly optimized for their instrument. An important consideration is that morphological operations are aimed at detecting RFI that is below the noise, therefore invisible to scientists that manually classify RFI. In the comparisons done in \citealt{nn-rfi-detection-2020}, samples detected by the morphological operator would all be counted as false positives, whereas this operator has been shown to improve the final science results \citep{offringa-2012-scale-invariant-rank-operator}. It can therefore not yet be stated that, based on accuracy, machine learning methods are outperforming traditional based methods. Rather, it is clear that both methods are competitive and are accurate enough to largely mitigate the problem of interference in radio data.

There are differences in the computational performance though. In \citet{xiao-2022}, machine learning methods flag a one-hour FAST observation of 67 GB in 61\% of the observing time using 8 computing nodes \citep{xiao-2022}. This amounts to a single-node computational performance of 14 GB/hour. On the other hand, the single-node performance of the \textsc{aoflagger} approach listed in \S\ref{sec:speed} is 370 MB/s, or 1.3 TB/hour, and \textsc{aoflagger} is therefore almost two orders of magnitude faster. While the performance of the computing nodes used for the computational performance analyses may differ somewhat, and it is therefore not a direct comparison, it is evident that the \textsc{aoflagger} approach is significantly faster. In \citet{sun-2022}, authors compare the run-time of \textsc{aoflagger} to their convolutional neural network (CNN) approach and find that \textsc{aoflagger} is two to four times faster. However, the authors measured the total run-time of the aoflagger executable, which would include disk access, start-up overhead and time spent in the \textsc{casacore} library to transfer the measurement set data. Because the flagging speed is near the disk access speed, this overhead can be substantial. A better benchmark is possible by using the C++ or Python API of \textsc{aoflagger} directly. On their \texttt{Sim\_RFI-1} dataset, they reach an \textsc{aoflagger} speed of 250 GB/hour, while in this work, with a more advanced strategy, we reach 1.3 TB/hour on similar hardware. Their CNN method reaches a speed of 145 GB/hour, which is an order of magnitude faster than what is reached by \citet{xiao-2022}, but is an order of magnitude below what we reach with our \textsc{aoflagger} approach.

\section{Discussion \& conclusions} \label{sec:discussion-conclusions}

We have described and demonstrated an automated RFI detection strategy aimed at flagging Apertif data. Our detection strategy implements novel \textsc{sumthreshold} and \textsc{sir}-operator algorithms that take prior information about invalid data into account. It also avoids the flagging of HI emission, works on auto-correlations, corrects the sub-band band-pass and contains some further parameter optimizations for Apertif. The change from the AOFlagger \texttt{XML} strategies towards fully scripted strategies provides flexibility that made these changes quite easy to implement and supports flexibility during experimentation. Besides making the process easier and faster, an automated RFI detection strategy also makes the results reproducible, compared to when RFI is flagged manually, and it allows reducing the data size by averaging early on in the data reduction processing.

We expect that our RFI detection strategy will work for data from other instruments, in particular those with a frequency coverage comparable to Apertif, such as MeerKAT, ASKAP, JVLA and future SKA-mid observations around 1.0 -- 1.5 GHz. Different bands might require some changes to the strategy parameters, but should be able to reuse a large part of the approach.

While machine learning techniques may compete with the accuracy of AOFlagger, they do not compete with its speed. Moreover, we have shown it is possible to add new features to AOFlagger, such as avoiding the 21-cm HI signal, accurate detection in the presence of invalid data and flagging of auto-correlations. None of the current available machine learning techniques support these scenarios. Most parameters, such as the sensitivity towards broadband and line RFI, or the expected smoothness of the data, are intuitive and easy to tweak for science cases that e.g. require that transients do not get flagged, or that require a difference balance between taking out all visible RFI on one hand, and keeping as much data available for further processing on the other hand. This will be challenging, if at all possible, to implement in a machine learning framework.

In this work, we have not made use of the multi-beaming capabilities of Apertif: beam are flagged independently. While some first-order testing indicates that using data integrated over all beams does not improve flagging accuracy, it can be expected that RFI does correlate somewhat over beams. A strategy where the integrated data is searched for RFI, and where this is used as additional input for the flagging of individual beams, might be effective for detecting RFI that is below the noise for a single beam.

\begin{acknowledgements}
This work makes use of data from the Apertif system installed at the Westerbork Synthesis Radio Telescope owned by ASTRON. ASTRON, the Netherlands Institute for Radio Astronomy, is an institute of the Dutch Research Council (de Nederlandse Organisatie voor Wetenschappelijk Onderzoek, NWO).
BA acknowledges funding from the German Science Foundation DFG, within the Collaborative Research Center SFB1491 ''Cosmic Interacting Matters - From Source to Signal''.
EAKA is supported by the WISE research programme, which is financed by NWO.
JMvdH and KMH, acknowledge funding from the European Research Council under the European Union’s Seventh
Framework Programme (FP/2007-2013)/ERC Grant Agreement No. 291531 (‘HIStoryNU’).
JvL, YM and LCO acknowledge funding from the European Research Council under the European Union's Seventh Framework Programme (FP/2007-2013)/ERC Grant Agreement No. 617199 (‘ALERT’; PI: JvL).
KMH further acknowledges financial support from the State Agency for Research of the Spanish Ministry of Science, Innovation and Universities through the ``Center of Excellence Severo Ochoa'' awarded to the Instituto de Astrof\'isica de Andaluc\'ia (SEV-2017-0709) from the coordination of the participation in SKA-SPAIN, funded by the Ministry of Science and innovation (MICIN) and grant RTI2018-096228-B-C31 (MCIU/AEI/FEDER,UE).
JvL further acknowledges funding from Vici research programme `ARGO' with project number 639.043.815, financed by NWO.
DV acknowledges support from the Netherlands eScience Center (NLeSC) under grant ASDI.15.406.
\end{acknowledgements}

% To make Dutch ``tussenvoegsels'' work correctly in Latex such as ``de Bruyn'', I use this command-6
% In bibliography, it should be written with lowercase, as in ``de Bruyn'' and
% the prefix should be ignored for the ordering.
\DeclareRobustCommand{\TUSSEN}[3]{#3}
\bibliographystyle{aa}
\bibliography{apertif-rfi-bibliography}

\end{document}